# Power Control for Two User Cooperative OFDMA Channels

Sezi Bakım Onur Kaya

*sezi.bakim@isik.edu.tr* *onurkaya@isikun.edu.tr*

**Abstract**

For a two user cooperative orthogonal frequency division multiple access (OFDMA) system with full channel state information (CSI), we obtain the optimal power allocation (PA) policies which maximize the rate region achievable by a channel adaptive implementation of inter-subchannel block Markov superposition encoding (BMSE), used in conjunction with backwards decoding. We provide the optimality conditions that need to be satisfied by the powers associated with the users' codewords and derive the closed form expressions for the optimal powers. We propose two algorithms that can be used to optimize the powers to achieve any desired rate pair on the rate region boundary: a projected subgradient algorithm, and an iterative waterfilling-like algorithm based on Karush-Kuhn-Tucker (KKT) conditions for optimality, which operates one user at a time and converges much faster. We observe that, utilization of power control to take advantage of the diversity offered by the cooperative OFDMA system, not only leads to a remarkable improvement in achievable rates, but also may help determine how the subchannels have to be instantaneously allocated to various tasks in cooperation.

## I. Introduction

The ability of OFDMA to cope with both intersymbol and interuser interference, combined with its low complexity of implementation, have made it a popular choice for the next generation wireless networks. As a result, the problem of resource allocation in OFDMA systems was studied extensively in the literature. One example is [2], where it was proved that in an OFDMA uplink system, allocating subcarriers to the users with the maximum marginal rate is a necessary condition for maximizing the system throughput. A similar problem was solved in [3] using KKT conditions, by optimizing a utility function which was assumed to be a function of the rates. In [4], a low-complexity algorithm for subcarrier, power, and rate allocation for OFDMA was proposed, to maximize the sum rate under individual rate constraints to guarantee fairness. The downlink ergodic sum rate maximization problem was considered in [5], where the authors

This work was supported by The Scientific & Technological Research Council of Turkey, Grant 108E208. A preliminary version of this work [1] was presented at IEEE Globecom 2011, Multicell Cooperation Workshop. The authors are with the Department of Electrical and Electronics Engineering, Işık University, Istanbul, Turkey.

developed a linear complexity subcarrier and power allocation algorithm. These works, as well as many others on OFDMA, naturally assume orthogonal multiple access, thereby choosing to avoid interference. However, like all orthogonal transmission techniques, OFDMA incurs some rate penalty. Moreover, “interference” in wireless channels is in fact free side information, and not ignoring it opens up the possibility of user cooperation. Therefore, in this paper, we focus on resource allocation for a two user OFDMA channel, which allows for mutual cooperation among the users over each subchannel, each taking into account the available side information.

The overheard information in a typical wireless multiple access channel (MAC), is captured by modeling the system as a MAC with generalized feedback (MAC-GF) [6]. In [6], achievable rates for the MAC-GF were obtained based on BMSE and backwards decoding. In [7], these encoding and decoding techniques were applied to a Gaussian MAC in fading, and the resulting rate regions were characterized. In [8], PA policies that maximize the rates achievable by BMSE for the same model were obtained.

While the above works all deal with a scalar MAC-GF, some works on resource allocation for user cooperation in vector channels, specifically OFDMA, also exist. A cooperative OFDMA system where each user is allowed to transmit and receive at the same time, but necessarily on different subcarriers, was considered in [9]. Subcarrier and PA schemes for a time division duplex amplify and forward protocol were employed in [10] with the aim of maximizing system throughput and enhancing fairness in a cooperative OFDMA uplink system. Resource allocation and cooperative partner selection in cooperative OFDM networks was investigated with the objective of minimizing the overall power in [11]. In [12], power allocation for an OFDM based two-way relay channel using physical network coding is considered. However, these works consider either a one sided cooperation strategy, or a mutually cooperative strategy based on two parallel dedicated relay channels, or mutual cooperation based on a time division protocol.

In this paper, we consider a more general cooperative OFDMA model recently introduced in [13] instead. This model is based on parallel MAC-GFs, and does not make any prior assumptions about the way in which the subchannels are assigned to the users. We extend the two full-duplex cooperative encoding strategies, namely intra-subchannel cooperative encoding (IntraSCE) and inter-subchannel cooperative encoding (InterSCE) of [13], to a channel adaptive scenario. The main contributions are (i) the characterization of the long term achievable rate region for a two user cooperative OFDMA system with power control; (ii) the analytical derivation of the optimal PA policy that results in the best known achievable rate for the non-orthogonal mutually cooperative scenario; (iii) the development of two algorithms which obtain the optimal PA, and (iv) the evaluation of the achievable rate region under several scenarios, including limited CSI feedback. We first obtain the properties of the PA policy that maximizes the sum rate of the cooperative OFDMA system employing IntraSCE and InterSCE. Despite the complex re-encoding structure employed in InterSCE, the achievable rate region turns out to be of a relatively similar form to its scalar counterpart, and we are able to extend some properties of the optimal PA derived in [8] for a scalar cooperative MAC, to cooperative OFDMA. As a result, the weighted sum of rates, which can be used to obtain any point on the rate region boundary, becomes concave, and convex optimization techniques can be employed. We first propose a projected subgradient algorithm that converges to the optimum and maximizes the achievable rate region. Next, we derive the optimality conditions, and closed form expressions for optimum powers analytically. We are then able to propose an alternative efficient iterative algorithm with a much lower complexity, to obtain the rate points on the achievable rate region boundary. This algorithm works by solving the KKT optimality conditions iteratively over the users, to obtain the optimal powers. As a result, we demonstrate that by jointly exploiting the diversity provided by OFDMA's parallel subchannels, and the temporal diversity created by the time varying channel, we obtain very promising gains in achievable rates. More interestingly, we observe that the optimal PA may automatically dictate that some subchannels are assigned exclusively to certain users/tasks, depending on the instantaneous channel state, and that, even with limited CSI feedback from the receiver, the improvement in the rate region is still significant.

## II. System Model

We consider a two user full-duplex cooperative OFDMA system with $N$ subchannels, which is shown in Fig. 1, and is modeled by,

$$Y_0^{(i)} = h_{10}^{(i)} X_1^{(i)} + h_{20}^{(i)} X_2^{(i)} + Z_0^{(i)}, \tag{1}$$

$$Y_1^{(i)} = h_{21}^{(i)} X_2^{(i)} + Z_1^{(i)}, \tag{2}$$

$$Y_2^{(i)} = h_{12}^{(i)} X_1^{(i)} + Z_2^{(i)}, \tag{3}$$

where, for each subchannel $i \in \{1, \ldots, N\}$, $X_k^{(i)}$ is the symbol transmitted by node $k$, $Z_l^{(i)}$ is the zero-mean additive white Gaussian noise at node $l$, with variance ${\sigma_l^{(i)}}^2$; $h_{kl}^{(i)}$ is the fading coefficient between nodes $k$ and $l$, and $Y_l^{(i)}$ is the symbol received at node $l$; with $k \in \{1, 2\}$, $l \in \{0, 1, 2\}$ and $k \neq l$. Here, the receiver is denoted by $l = 0$. To simplify the notation throughout the paper, we define the normalized power-fading coefficients $s_{kl}^{(i)} = \frac{(h_{kl}^{(i)})^2}{{\sigma_l^{(i)}}^2}$, and the Gaussian capacity function $C(x) \triangleq \frac{1}{2}\log(1+x)$. For a real number $x$, we define $(x)^+ = \max(x, 0)$.

## III. Long-term Achievable Rates for Cooperative OFDMA

We first briefly review the channel non-adaptive IntraSCE and InterSCE strategies proposed in [13], which shall be extended to obtain our channel adaptive model and rate regions. Both mutually cooperative strategies are of decode and forward type, and rely on block Markov superposition encoding at the transmitters, and backward decoding at the receiver. The communication takes place in $B$ blocks. The message $w_k[b]$ of each user $k \in \{1, 2\}$ in block $b$ is divided into two submessages, $w_{k0}[b]$ and $w_{kj}[b]$, intended to be decoded at the receiver and cooperative partner $j \in \{1, 2\}$ respectively, which are further divided into $N$ submessages each,

$$w_{k0}[b] = \left\{ w_{k0}^{(1)}[b], ..., w_{k0}^{(N)}[b] \right\}, \quad w_{kj}[b] = \left\{ w_{kj}^{(1)}[b], ..., w_{kj}^{(N)}[b] \right\}, \tag{4}$$

to be transmitted over disjoint subchannels. In both IntraSCE and InterSCE, the transmitted codeword by each user $k$ over each subchannel $i$ in block $b \in \{1, \ldots, B\}$ is given by,

$$X_k^{(i)} = \sqrt{p_{k0}^{(i)}} X_{k0}^{(i)} + \sqrt{p_{kj}^{(i)}} X_{kj}^{(i)} + \sqrt{p_{U_k}^{(i)}} U_k^{(i)}. \tag{5}$$

Here, the component codewords $X_{k0}^{(i)}$, $X_{kj}^{(i)}$ and $U_k^{(i)}$ are all selected from codebooks which are randomly generated from unit Gaussian distributions. In a given block $b$, the task of $X_{k0}^{(i)}$ is to transmit fresh information $w_{k0}^{(i)}[b]$ directly intended for the receiver; while the codeword $X_{kj}^{(i)}$ is used for establishing common information, $w_{kj}^{(i)}[b]$, at the cooperating partner. User $j$ decodes $w_{kj}^{(i)}[b]$ at the end of block $b$ using $X_{kj}^{(i)}$, and treating $X_{k0}^{(i)}$ as noise. The difference of IntraSCE and InterSCE lies in the way $U_k^{(i)}$, which is the codeword used for conveying the previously established common information to the receiver, is mapped to the messages. In IntraSCE, $U_k^{(i)}$ is used to re-transmit the cooperative submessages, $w_{kj}^{(i)}[b-1]$ and $w_{jk}^{(i)}[b-1]$ received on subchannel $i$ in block $b-1$, to the destination, over the same sub-channel. However in the InterSCE strategy, after common information is established at the cooperating partner, the cooperative messages are re-partitioned, and $U_k^{(i)}$ may be used to transmit sub-messages received over other subchannels. Note that, since both users will know $w_{kj}[b-1]$ and $w_{jk}[b-1]$ at the end of block $b-1$, $U_k^{(i)}$ is commonly known to both users, and does not act as further interference while decoding $w_{kj}^{(i)}[b]$. The details of the achievability scheme for the channel non-adaptive case can be found in [13].

Note that, (5) does not utilize instantaneous CSI to adapt the instantaneous transmission powers. However, if we assume that the users and the receiver have full CSI of both the cooperative links and the direct link, the users can further adapt their transmitted symbols $X_k^{(i)}$ as a function of the joint fading state s, to maximize the long term (ergodic) achievable rates. In general, there are two ways to perform such channel adaptation: we can either use a variable power, variable rate codebook, as in [14], or we can use a single codebook, whose rate is supported by the channel in the long term, and perform the channel adaptation by simply multiplying entries from this codebook by channel adaptive powers, as in [15]. In this paper, we employ the latter approach, and propose a channel adaptive version of the encoding strategies in [13], where we scale each of the codewords in (5) by variable powers,

$$X_k^{(i)} = \sqrt{p_{k0}^{(i)}(\mathbf{s})}X_{k0}^{(i)} + \sqrt{p_{kj}^{(i)}(\mathbf{s})}X_{kj}^{(i)} + \sqrt{p_{U_k}^{(i)}(\mathbf{s})}U_k^{(i)}, \tag{6}$$

where $k, j \in \{1, 2\}$, $k \neq j$, $i = 1, \cdots, N$. The powers are subject to the average power constraints,

$$\sum_{i=1}^{N} E\left[p_{k0}^{(i)}(\mathbf{s}) + p_{kj}^{(i)}(\mathbf{s}) + p_{U_k}^{(i)}(\mathbf{s})\right] \triangleq \sum_{i=1}^{N} E\left[p_k^{(i)}(\mathbf{s})\right] \leq \bar{p}_k. \tag{7}$$

The achievable rate regions for power controlled IntraSCE and InterSCE are obtained by extending [13, Corollary 1] and [13, Corollary 2] respectively, using the new the channel adaptive encoding defined in (6). The resulting achievable rate region for IntraSCE with power control is given by the closure of the convex hull of all rate pairs $(R_1, R_2)$ satisfying

$$R_1 < \sum_{i=1}^{N} \min\left\{ E\left[ C\left( \frac{s_{12}^{(i)} p_{12}^{(i)}(\mathbf{s})}{s_{12}^{(i)} p_{10}^{(i)}(\mathbf{s}) + 1} \right) + C\left( s_{10}^{(i)} p_{10}^{(i)}(\mathbf{s}) \right) \right], \right.$$

$$E\left[C\left(s_{10}^{(i)}p_1^{(i)}(\mathbf{s})+s_{20}^{(i)}p_2^{(i)}(\mathbf{s})+2\sqrt{s_{10}^{(i)}s_{20}^{(i)}p_{U_1}^{(i)}(\mathbf{s})p_{U_2}^{(i)}(\mathbf{s})}\right)\right]\Bigg\}, \quad (8)$$

$$R_2<\sum_{i=1}^{N}\min\left\{E\left[C\left(\frac{s_{21}^{(i)}p_{21}^{(i)}(\mathbf{s})}{s_{21}^{(i)}p_{20}^{(i)}(\mathbf{s})+1}\right)+C\left(s_{20}^{(i)}p_{20}^{(i)}(\mathbf{s})\right)\right],\right.$$

$$E\left[C\left(s_{10}^{(i)}p_1^{(i)}(\mathbf{s})+s_{20}^{(i)}p_2^{(i)}(\mathbf{s})+2\sqrt{s_{10}^{(i)}s_{20}^{(i)}p_{U_1}^{(i)}(\mathbf{s})p_{U_2}^{(i)}(\mathbf{s})}\right)\right]\Bigg\}, \quad (9)$$

$$R_1+R_2<\sum_{i=1}^{N}\min\left\{E\left[C\left(s_{10}^{(i)}p_1^{(i)}(\mathbf{s})+s_{20}^{(i)}p_2^{(i)}(\mathbf{s})+2\sqrt{s_{10}^{(i)}s_{20}^{(i)}p_{U_1}^{(i)}(\mathbf{s})p_{U_2}^{(i)}(\mathbf{s})}\right)\right],\right.$$

$$E\left[C\left(\frac{s_{12}^{(i)}p_{12}^{(i)}(\mathbf{s})}{s_{12}^{(i)}p_{10}^{(i)}(\mathbf{s})+1}\right)+C\left(\frac{s_{21}^{(i)}p_{21}^{(i)}(\mathbf{s})}{s_{21}^{(i)}p_{20}^{(i)}(\mathbf{s})+1}\right)+C\left(s_{10}^{(i)}p_{10}^{(i)}(\mathbf{s})+s_{20}^{(i)}p_{20}^{(i)}(\mathbf{s})\right)\right]\Bigg\}, \quad (10)$$

and the achievable rate region for InterSCE with power control is given by the closure of the convex hull of all rate pairs $(R_1, R_2)$ satisfying

$$R_1<\sum_{i=1}^{N}E\left[C\left(\frac{s_{12}^{(i)}p_{12}^{(i)}(\mathbf{s})}{s_{12}^{(i)}p_{10}^{(i)}(\mathbf{s})+1}\right)+C\left(s_{10}^{(i)}p_{10}^{(i)}(\mathbf{s})\right)\right], \quad (11)$$

$$R_2<\sum_{i=1}^{N}E\left[C\left(\frac{s_{21}^{(i)}p_{21}^{(i)}(\mathbf{s})}{s_{21}^{(i)}p_{20}^{(i)}(\mathbf{s})+1}\right)+C\left(s_{20}^{(i)}p_{20}^{(i)}(\mathbf{s})\right)\right], \quad (12)$$

$$R_1+R_2<\min\left\{\sum_{i=1}^{N}E\left[C\left(s_{10}^{(i)}p_1^{(i)}(\mathbf{s})+s_{20}^{(i)}p_2^{(i)}(\mathbf{s})+2\sqrt{s_{10}^{(i)}s_{20}^{(i)}p_{U_1}^{(i)}(\mathbf{s})p_{U_2}^{(i)}(\mathbf{s})}\right)\right],\right.$$

$$\sum_{i=1}^{N}E\left[C\left(\frac{s_{12}^{(i)}p_{12}^{(i)}(\mathbf{s})}{s_{12}^{(i)}p_{10}^{(i)}(\mathbf{s})+1}\right)+C\left(\frac{s_{21}^{(i)}p_{21}^{(i)}(\mathbf{s})}{s_{21}^{(i)}p_{20}^{(i)}(\mathbf{s})+1}\right)+C\left(s_{10}^{(i)}p_{10}^{(i)}(\mathbf{s})+s_{20}^{(i)}p_{20}^{(i)}(\mathbf{s})\right)\right]\Bigg\}, \quad (13)$$

where the convex hulls are taken over all valid PA policies. In the next section, we obtain the PA policies which achieve the rate tuples on the rate region boundary. To do this, we first derive a simplifying property of optimal PA for both cooperative encoding strategies, and then we focus on InterSCE that provides superior achievable rates.

## IV. Channel Adaptive Power Allocation

If we set $N = 1$ in (8)-(10) or (11)-(13), the problem reduces to a scalar cooperative MAC. In [8], it was shown for this scalar case that, based on the instantaneous channel state, the optimal PA dictates that each user either sends cooperative information, or fresh information, but not both. Although in OFDMA, there is a sum power constraint over the subchannels, and one would expect the PA over each subchannel to be dependent on the powers assigned to the other subchannels, we show that many properties of the optimal PA for the proposed cooperative OFDMA system remain surprisingly parallel to those in the scalar case [8], and the codewords that should be used over each subchannel are determined solely by the instantaneous fading coefficients over that particular subchannel, as stated in the following lemma:

*Lemma 1:* The PA policy that maximizes the sum rate of a cooperative OFDMA system using IntraSCE and InterSCE should satisfy;

1) $p_{10}^{(i)*}(\mathbf{s}) = p_{20}^{(i)*}(\mathbf{s}) = 0$, if $\mathbf{s} \in \mathcal{S}_1$,
2) $p_{10}^{(i)*}(\mathbf{s}) = p_{21}^{(i)*}(\mathbf{s}) = 0$, if $\mathbf{s} \in \mathcal{S}_2$,
3) $p_{12}^{(i)*}(\mathbf{s}) = p_{20}^{(i)*}(\mathbf{s}) = 0$, if $\mathbf{s} \in \mathcal{S}_3$,
4) $p_{12}^{(i)*}(\mathbf{s}) = p_{21}^{(i)*}(\mathbf{s}) = 0$ or $p_{10}^{(i)*}(\mathbf{s}) = p_{21}^{(i)*}(\mathbf{s}) = 0$ or $p_{12}^{(i)*}(\mathbf{s}) = p_{20}^{(i)*}(\mathbf{s}) = 0$, if $\mathbf{s} \in \mathcal{S}_4$,

where $\mathcal{S}_1 = \{\mathbf{s} : \ s_{12}^{(i)} > s_{10}^{(i)}, \ s_{21}^{(i)} > s_{20}^{(i)}\}$, $\mathcal{S}_2 = \{\mathbf{s} : \ s_{12}^{(i)} > s_{10}^{(i)}, \ s_{21}^{(i)} \leq s_{20}^{(i)}\}$, $\mathcal{S}_3 = \{\mathbf{s} : \ s_{12}^{(i)} \leq s_{10}^{(i)}, \ s_{21}^{(i)} > s_{20}^{(i)}\}$, $\mathcal{S}_4 = \{\mathbf{s} : \ s_{12}^{(i)} \leq s_{10}^{(i)}, \ s_{21}^{(i)} \leq s_{20}^{(i)}\}$.

*Proof:* Assume that we know the total optimal power $p_k^{(i)*}(\mathbf{s})$, allocated to each subchannel $i$ at each channel state $\mathbf{s}$. Then, for IntraSCE, the sum rate (10) is maximized if each term in the summation is maximized. Since the total power allocated to each term is fixed, we have $N$ independent optimization problems, and by [8, Proposition 1] the result follows. For InterSCE, the sum rate (13) is maximized if each argument of the minimum operation is maximized. The first argument of (13) is insensitive to the choice of $p_{k0}^{(i)*}(\mathbf{s})$ or $p_{kj}^{(i)*}(\mathbf{s})$, as long as their sum is fixed; whereas the second argument is maximized if we separately maximize its summands for each $i$. The result follows by noting that this is also equivalent to $N$ independent optimization problems, each yielding a scalar case, and [8, Proposition 1] holds, giving the desired result. ■

An important observation is that, setting two of the powers equal to zero as suggested by Lemma 1, is also optimal for the entire rate region maximization, as the right hand sides of all three constraints, for both policies, are maximized by choosing the powers according to Lemma 1.[1] Therefore, from now on we focus only on policies that satisfy Lemma 1.

Note that, the bounds (11), (12) and (13) on $R_1$, $R_2$ and $R_1 + R_2$ respectively for InterSCE are looser than the corresponding bounds (8), (9) and (10) for IntraSCE, as the minimum operations in (8), (9) are removed, and the minimum in (10) is taken outside the summation, to obtain (11), (12) and (13). As a result, the achievable rate region of InterSCE contains that of IntraSCE. Hence, it is sufficient to limit our focus on the InterSCE strategy, which results in a uniformly better rate region. Then, it is easy to check that the rate constraints in (11)-(13) now become concave in the power vector $\mathbf{p}(\mathbf{s}) = [p_{10}^{(i)*}(\mathbf{s}), p_{12}^{(i)*}(\mathbf{s}), p_{U_1}^{(i)*}(\mathbf{s}), p_{20}^{(i)*}(\mathbf{s}), p_{21}^{(i)*}(\mathbf{s}), p_{U_2}^{(i)*}(\mathbf{s}),\ i = 1, \ldots, N]$, lending themselves to well known techniques in convex optimization, which we discuss in the next sections.

[1] We choose the first option for $\mathbf{s} \in \mathcal{S}_4$, which may cause a slight deviation from optimality for the sum rate. However, this case rarely occurs in practice, and this suboptimality can be ignored, as it has been done in [8].

### *A. Achievable Rate Maximization Using Projected Subgradients*

Since all bounds of the achievable rate region are concave in powers, so is any weighted sum $\mu_1 R_1 + \mu_2 R_2$ at the corners. Moreover, it is easy to show that the rate region is strictly convex [8], [15]. Therefore, we can obtain points on the rate region boundary by maximizing $R_{\boldsymbol{\mu}} = \mu_1 R_1 + \mu_2 R_2$, where $\{R_1, R_2\}$ is the corner of the pentagon obtained for a given PA policy, defined by (11)-(13). Assuming $\mu_1 > \mu_2$ without loss of generality, and employing Lemma 1 to simplify (11)-(13), the optimization problem can be stated as:

$$
\begin{aligned}
\max_{\mathbf{p}(\mathbf{s})} \Bigg( & (\mu_1 - \mu_2) \sum_{i=1}^{N} \left( E_{\mathcal{S}_1,\mathcal{S}_2} \left[ C\left( s_{12}^{(i)} p_{12}^{(i)}(\mathbf{s}) \right) \right] + E_{\mathcal{S}_3,\mathcal{S}_4} \left[ C\left( s_{10}^{(i)} p_{10}^{(i)}(\mathbf{s}) \right) \right] \right) \\
& + \mu_2 \min \Bigg\{ \sum_{i=1}^{N} E \left[ C\left( s_{10}^{(i)} p_1^{(i)}(\mathbf{s}) + s_{20}^{(i)} p_2^{(i)}(\mathbf{s}) + 2\sqrt{s_{10}^{(i)} s_{20}^{(i)} p_{U_1}^{(i)}(\mathbf{s}) p_{U_2}^{(i)}(\mathbf{s})} \right) \right], \\
& \sum_{i=1}^{N} \Bigg( E_{\mathcal{S}_1} \left[ C\left( s_{12}^{(i)} p_{12}^{(i)}(\mathbf{s}) \right) + C\left( s_{21}^{(i)} p_{21}^{(i)}(\mathbf{s}) \right) \right] + E_{\mathcal{S}_2} \left[ C\left( s_{12}^{(i)} p_{12}^{(i)}(\mathbf{s}) \right) + C\left( s_{20}^{(i)} p_{20}^{(i)}(\mathbf{s}) \right) \right] \\
& \quad + E_{\mathcal{S}_3} \left[ C\left( s_{10}^{(i)} p_{10}^{(i)}(\mathbf{s}) \right) + C\left( s_{21}^{(i)} p_{21}^{(i)}(\mathbf{s}) \right) \right] + E_{\mathcal{S}_4} \left[ C\left( s_{10}^{(i)} p_{10}^{(i)}(\mathbf{s}) + s_{20}^{(i)} p_{20}^{(i)}(\mathbf{s}) \right) \right] \Bigg) \Bigg\} \Bigg),
\end{aligned}
\tag{14}
$$

$$
\begin{aligned}
\text{s.t. } & \sum_{i=1}^{N} E \left[ p_{k0}^{(i)}(\mathbf{s}) + p_{kj}^{(i)}(\mathbf{s}) + p_{U_k}^{(i)}(\mathbf{s}) \right] \leq \bar{p}_k, \\
& p_{k0}^{(i)}(\mathbf{s}), p_{kj}^{(i)}(\mathbf{s}), p_{U_k}^{(i)}(\mathbf{s}) \geq 0,\ k, j \in \{1, 2\},\ k \neq j,
\end{aligned}
$$

where $E_{\mathcal{S}_d}$ denotes the expectation over $\mathbf{s} \in \mathcal{S}_d$, $d = 1, 2, 3, 4$.

Due to the minimum operation in (14), the gradient of the objective function does not exist everywhere. In particular, there are two gradient vectors, depending on which argument of the minimum in (14) is active. Yet, these vectors may be viewed instead as subgradients, which makes it possible to employ the method of projected subgradients, for power optimization. Due to the convex nature of our constraints, this method is guaranteed to converge to the global optimum [16], with a diminishing stepsize normalized by the norm of the subgradient.

Since the calculation of the subgradients requires rather tedious formulas which give little insight, we will directly provide some examples of the achievable rate region, and the resulting PA policy, based on simulations in Section V instead. The major drawbacks of the subgradient algorithm are its slow rate of convergence, and complexity. As the number of subchannels increase, so does the size of the vector of power variables, making the process of computing the subgradients, and the projection operations formidable. Hence, we next obtain analytical expressions for the weighted sum-rate optimal power control, and propose an alternative iterative algorithm which converges much faster than the subgradient algorithm.

*B. Iterative Achievable Rate Maximization Based on KKT Conditions*

The optimization problem (14), can be stated in an equivalent differentiable form

$$\max_{\mathbf{p}(\mathbf{s})} \quad R_\mu$$

$$\begin{aligned} s.t.\ R_\mu \leq (\mu_1-\mu_2)\sum_{i=1}^{N}\Big(E_{S_1,S_2}\Big[C\Big(s_{12}^{(i)}p_{12}^{(i)}(\mathbf{s})\Big)\Big]+E_{S_3,S_4}\Big[C\Big(s_{10}^{(i)}p_{10}^{(i)}(\mathbf{s})\Big)\Big]\Big) \\ +\mu_2\sum_{i=1}^{N}E\left[C\left(s_{10}^{(i)}p_1^{(i)}(\mathbf{s})+s_{20}^{(i)}p_2^{(i)}(\mathbf{s})+2\sqrt{s_{10}^{(i)}s_{20}^{(i)}p_{U_1}^{(i)}(\mathbf{s})p_{U_2}^{(i)}(\mathbf{s})}\right)\right], \end{aligned} \tag{15}$$

$$\begin{aligned} R_\mu \leq (\mu_1-\mu_2)\sum_{i=1}^{N}\Big(E_{S_1,S_2}\Big[C\Big(s_{12}^{(i)}p_{12}^{(i)}(\mathbf{s})\Big)\Big]+E_{S_3,S_4}\Big[C\Big(s_{10}^{(i)}p_{10}^{(i)}(\mathbf{s})\Big)\Big]\Big) \\ +\mu_2\sum_{i=1}^{N}\Big(E_{S_1}\Big[C\Big(s_{12}^{(i)}p_{12}^{(i)}(\mathbf{s})\Big)+C\Big(s_{21}^{(i)}p_{21}^{(i)}(\mathbf{s})\Big)\Big]+E_{S_2}\Big[C\Big(s_{12}^{(i)}p_{12}^{(i)}(\mathbf{s})\Big)+C\Big(s_{20}^{(i)}p_{20}^{(i)}(\mathbf{s})\Big)\Big] \\ +E_{S_3}\Big[C\Big(s_{10}^{(i)}p_{10}^{(i)}(\mathbf{s})\Big)+C\Big(s_{21}^{(i)}p_{21}^{(i)}(\mathbf{s})\Big)\Big]+E_{S_4}\Big[C\Big(s_{10}^{(i)}p_{10}^{(i)}(\mathbf{s})+s_{20}^{(i)}p_{20}^{(i)}(\mathbf{s})\Big)\Big]\Big), \end{aligned} \tag{16}$$

$$\sum_{i=1}^{N}\Big(E_{S_3,S_4}\Big[p_{10}^{(i)}(\mathbf{s})\Big]+E_{S_1,S_2}\Big[p_{12}^{(i)}(\mathbf{s})\Big]+E\Big[p_{U_1}^{(i)}(\mathbf{s})\Big]\Big)\leq \bar{p}_1, \tag{17}$$

$$\sum_{i=1}^{N}\Big(E_{S_2,S_4}\Big[p_{20}^{(i)}(\mathbf{s})\Big]+E_{S_1,S_3}\Big[p_{21}^{(i)}(\mathbf{s})\Big]+E\Big[p_{U_2}^{(i)}(\mathbf{s})\Big]\Big)\leq \bar{p}_2, \tag{18}$$

$$p_{10}^{(i)}(\mathbf{s}),p_{12}^{(i)}(\mathbf{s}),p_{U_1}^{(i)}(\mathbf{s}),p_{20}^{(i)}(\mathbf{s}),p_{21}^{(i)}(\mathbf{s}),p_{U_2}^{(i)}(\mathbf{s})\geq 0, \quad \forall \mathbf{s}. \tag{19}$$

Note that, due to the concavity of the logarithm, (15)-(19) is a convex optimization problem, with differentiable constraints, and hence the KKT conditions are necessary and sufficient for optimality. Assigning the Lagrange multipliers $\gamma_1, \gamma_2, \lambda_1, \lambda_2$ to the constraints (15)-(18), and $\epsilon_t^{(i)}(\mathbf{s})$, $t = 1, ..., 6$, to the positivity constraints (19), we obtain the conditions for optimality, given in the following lemma.

*Lemma 2:* Define the variable $A^{(i)}$, $i = 1, \cdots, N$; and the indices $m$, $n$ as follows:

$$A^{(i)} = 1 + s_{10}^{(i)} p_1^{(i)}(\mathbf{s}) + s_{20}^{(i)} p_2^{(i)}(\mathbf{s}) + 2\sqrt{s_{10}^{(i)} s_{20}^{(i)} p_{U_1}^{(i)}(\mathbf{s}) p_{U_2}^{(i)}(\mathbf{s})}, \tag{20}$$

$$m = \begin{cases} 0, & \text{if } \mathbf{s} \in \mathcal{S}_3 \cup \mathcal{S}_4 \\ 2, & \text{if } \mathbf{s} \in \mathcal{S}_1 \cup \mathcal{S}_2 \end{cases}, \quad n = \begin{cases} 0, & \text{if } \mathbf{s} \in \mathcal{S}_2 \cup \mathcal{S}_4 \\ 1, & \text{if } \mathbf{s} \in \mathcal{S}_1 \cup \mathcal{S}_3 \end{cases}. \tag{21}$$

A power allocation policy $p_{1m}^{(i)}(\mathbf{s})$, $p_{2n}^{(i)}(\mathbf{s})$, $p_{U_1}^{(i)}(\mathbf{s})$, $p_{U_2}^{(i)}(\mathbf{s})$ is optimal for the problem (15)-(19), if and only if it satisfies, for s $\in \mathcal{S}_1 \cup \mathcal{S}_2 \cup \mathcal{S}_3 \triangleq \mathcal{S}_4^c$,

$$(\mu_1 - \mu_2 + \gamma_1 \mu_2) \frac{s_{1m}^{(i)}}{1 + s_{1m}^{(i)} p_{1m}^{(i)}(\mathbf{s})} + \gamma_2 \mu_2 \frac{s_{10}^{(i)}}{A^{(i)}} \leq \lambda_1, \tag{22}$$

$$\gamma_1 \mu_2 \frac{s_{2n}^{(i)}}{1 + s_{2n}^{(i)} p_{2n}^{(i)}(\mathbf{s})} + \gamma_2 \mu_2 \frac{s_{20}^{(i)}}{A^{(i)}} \leq \lambda_2, \tag{23}$$

$$\gamma_2 \mu_2 \frac{\sqrt{s_{k0}^{(i)} s_{j0}^{(i)} p_{U_j}^{(i)}(\mathbf{s})} + s_{k0}^{(i)} \sqrt{p_{U_k}^{(i)}(\mathbf{s})}}{A^{(i)} \sqrt{p_{U_k}^{(i)}(\mathbf{s})}} \leq \lambda_k, \quad k \in \{1, 2\} \tag{24}$$

and for s $\in \mathcal{S}_4$,

$$(\mu_1 - \mu_2) \frac{s_{10}^{(i)}}{1 + s_{10}^{(i)} p_{1m}^{(i)}(\mathbf{s})} + \gamma_1 \mu_2 \frac{s_{10}^{(i)}}{1 + s_{10}^{(i)} p_{1m}^{(i)}(\mathbf{s}) + s_{20}^{(i)} p_{2n}^{(i)}(\mathbf{s})} + \gamma_2 \mu_2 \frac{s_{10}^{(i)}}{A^{(i)}} \leq \lambda_1, \tag{25}$$

$$\gamma_1 \mu_2 \frac{s_{20}^{(i)}}{1 + s_{20}^{(i)} p_{2n}^{(i)}(\mathbf{s}) + s_{10}^{(i)} p_{1m}^{(i)}(\mathbf{s})} + \gamma_2 \mu_2 \frac{s_{20}^{(i)}}{A^{(i)}} \leq \lambda_2, \tag{26}$$

$$\gamma_2 \mu_2 \frac{\sqrt{s_{k0}^{(i)} s_{j0}^{(i)} p_{U_j}^{(i)}(\mathbf{s})} + s_{k0}^{(i)} \sqrt{p_{U_k}^{(i)}(\mathbf{s})}}{A^{(i)} \sqrt{p_{U_k}^{(i)}(\mathbf{s})}} \leq \lambda_k, \quad k \in \{1, 2\} \tag{27}$$

where the Lagrange multipliers $\gamma_1$, $\gamma_2 = 1 - \gamma_1$, $\lambda_1$ and $\lambda_2$ are selected so that the constraints (15)-(18) are satisfied with equality. Each of the constraints (22), (23) and (24) (correspondingly (25), (26) and (27) when s $\in \mathcal{S}_4$) are satisfied with equality if and only if the respective power levels, $p_{1m}^{(i)}(\mathbf{s})$, $p_{2n}^{(i)}(\mathbf{s})$ or $p_{U_k}^{(i)}(\mathbf{s})$ are positive.

*Proof:* See Appendix. ■

The optimality conditions given in Lemma 2 for each power component are heavily coupled, thereby making the computation of the optimal PA policy seemingly difficult. Yet, in the following theorem, we show that, after some non-trivial observations, the coupling among the constraints is partially removed, and as a result, we are able to provide closed form expressions for the optimal power levels.

*Theorem 1:* For a cooperative OFDMA system employing InterSCE, the optimal PA, $p_{1m}^{(i)}(\mathbf{s})$, $p_{2n}^{(i)}(\mathbf{s})$, $p_{U_1}^{(i)}(\mathbf{s})$, $p_{U_2}^{(i)}(\mathbf{s})$, that solves (15)-(19) is given by

$$p_{U_k}^{(i)}(\mathbf{s}) = s_{k0}^{(i)} \frac{\frac{\mu_2(1-\gamma_1)}{\lambda_k}\left(s_{k0}^{(i)} + \frac{\lambda_k}{\lambda_j} s_{j0}^{(i)}\right) - \left(1 + s_{10}^{(i)} p_{1m}^{(i)} + s_{20}^{(i)} p_{2n}^{(i)}\right)}{\left(s_{k0}^{(i)} + \frac{\lambda_k}{\lambda_j} s_{j0}^{(i)}\right)^2}, \tag{28}$$

$$p_{1m}^{(i)}(\mathbf{s}) = \begin{cases} \left(\frac{(\mu_1-\mu_2+\gamma_1\mu_2)\left(\lambda_2 s_{10}^{(i)}+\lambda_1 s_{20}^{(i)}\right)}{\lambda_1^2 s_{20}^{(i)}} - \frac{1}{s_{1m}^{(i)}}\right)^+, & \text{if } \mathbf{s} \in \mathcal{S}_4^c \quad \text{(29a)} \\ f\Bigg({s_{10}^{(i)}}^2, \frac{(\lambda_1 s_{20}^{(i)}+\lambda_2 s_{10}^{(i)})(\mu_1-\mu_2+\gamma_1\mu_2){s_{10}^{(i)}}^2-{\lambda_1}^2 s_{20}^{(i)}(2s_{10}^{(i)}+s_{10}^{(i)}s_{20}^{(i)}p_{2n}^{(i)}(\mathbf{s}))}{-{\lambda_1}^2 s_{20}^{(i)}}, & \\ \frac{(\lambda_1 s_{20}^{(i)}+\lambda_2 s_{10}^{(i)})\left[(\mu_1-\mu_2+\gamma_1\mu_2)+(\mu_1-\mu_2)s_{20}^{(i)}p_{2n}^{(i)}(\mathbf{s})\right]s_{10}^{(i)}-{\lambda_1}^2 s_{20}^{(i)}(1+s_{20}^{(i)}p_{2n}^{(i)}(\mathbf{s}))}{-{\lambda_1}^2 s_{20}^{(i)}}\Bigg), & \text{if } \mathbf{s} \in \mathcal{S}_4 \quad \text{(29b)} \end{cases}$$

$$p_{2n}^{(i)}(\mathbf{s}) = \begin{cases} \left(\frac{\gamma_1\mu_2\left(\lambda_2 s_{10}^{(i)}+\lambda_1 s_{20}^{(i)}\right)}{\lambda_2^2 s_{10}^{(i)}} - \frac{1}{s_{2n}^{(i)}}\right)^+, & \text{if } \mathbf{s} \in \mathcal{S}_4^c \quad \text{(30a)} \\ \left(\frac{\gamma_1\mu_2\left(\lambda_2 s_{10}^{(i)}+\lambda_1 s_{20}^{(i)}\right)}{\lambda_2^2 s_{10}^{(i)}} - \frac{1}{s_{20}^{(i)}} - \frac{s_{10}^{(i)}}{s_{20}^{(i)}} p_{1m}^{(i)}(\mathbf{s})\right)^+, & \text{if } \mathbf{s} \in \mathcal{S}_4 \quad \text{(30b)} \end{cases}$$

if the powers obtained from (28) are positive, i.e., $p_{U_k}^{(i)}(\mathbf{s}) > 0$; and

$$p_{U_k}^{(i)}(\mathbf{s}) = 0, \tag{31}$$

$$p_{1m}^{(i)}(\mathbf{s}) = \begin{cases} f\Big(\lambda_1 s_{10}^{(i)} s_{1m}^{(i)}, -\mu_1 s_{10}^{(i)} s_{1m}^{(i)} + \lambda_1(s_{10}^{(i)} + s_{1m}^{(i)} + s_{1m}^{(i)} s_{20}^{(i)} p_{2n}^{(i)}(\mathbf{s})), \lambda_1(1 + s_{20}^{(i)} p_{2n}^{(i)}(\mathbf{s})) & \\ -(\mu_1 - \mu_2 + \gamma_1\mu_2)\, s_{1m}^{(i)}(1 + s_{20}^{(i)} p_{2n}^{(i)}(\mathbf{s})) - \mu_2(1-\gamma_1) s_{10}^{(i)}\Big), & \text{if } \mathbf{s} \in \mathcal{S}_4^c \quad \text{(32a)} \\ f\Big(\lambda_1 {s_{10}^{(i)}}^2, -\mu_1 {s_{10}^{(i)}}^2 + \lambda_1 s_{10}^{(i)}(2 + s_{20}^{(i)} p_{2n}^{(i)}(\mathbf{s})), & \\ -\mu_1 s_{10}^{(i)} - (\mu_1 - \mu_2) s_{10}^{(i)} s_{20}^{(i)} p_{2n}^{(i)}(\mathbf{s}) + \lambda_1(1 + s_{20}^{(i)} p_{2n}^{(i)}(\mathbf{s}))\Big), & \text{if } \mathbf{s} \in \mathcal{S}_4 \quad \text{(32b)} \end{cases}$$

$$p_{2n}^{(i)}(\mathbf{s}) = \begin{cases} f\Big(\lambda_2 s_{20}^{(i)} s_{2n}^{(i)}, -\mu_2 s_{20}^{(i)} s_{2n}^{(i)} + \lambda_2(s_{20}^{(i)} + s_{2n}^{(i)} + s_{10}^{(i)} s_{2n}^{(i)} p_{1m}^{(i)}(\mathbf{s})), \lambda_2(1 + s_{10}^{(i)} p_{1m}^{(i)}(\mathbf{s})) \\ -\gamma_1 \mu_2 s_{2n}^{(i)}(1 + s_{10}^{(i)} p_{1m}^{(i)}(\mathbf{s})) - \mu_2(1-\gamma_1) s_{20}^{(i)}\Big), & \text{if } \mathbf{s} \in \mathcal{S}_4^c \quad \text{(33a)} \\ \left(\frac{\mu_2}{\lambda_2} - \frac{1}{s_{20}^{(i)}} - \frac{s_{10}^{(i)}}{s_{20}^{(i)}} p_{1m}^{(i)}(\mathbf{s})\right)^+, & \text{if } \mathbf{s} \in \mathcal{S}_4 \quad \text{(33b)} \end{cases}$$

otherwise, where $\gamma_1$, $\lambda_1$ and $\lambda_2$ are selected to satisfy the constraints (15)-(18) with equality, and the function $f(\cdot,\cdot,\cdot)$ is defined as $f(a,b,c) \triangleq \left(\frac{-b+\sqrt{b^2-4ac}}{2a}\right)^+$.

*Proof:* We start by noting that, to obtain coherent combining gain, the optimal cooperative powers $p_{U_k}^{(i)}(\mathbf{s})$, $k = 1, 2$, over a given subchannel and given channel state s, should either be both positive, or both zero. Let us first assume that both $p_{U_1}^{(i)}(\mathbf{s})$ and $p_{U_2}^{(i)}(\mathbf{s})$ are positive. Then, the constraints (24), (equivalently (27)) should be satisfied with equality, for $k = 1, 2$. Evaluating (24), (equivalently (27)) separately for $k = 1, 2$, and dividing the resulting equalities, we get

$$\frac{\sqrt{s_{20}^{(i)}}\sqrt{p_{U_2}^{(i)}(\mathbf{s})} + \sqrt{s_{10}^{(i)}}\sqrt{p_{U_1}^{(i)}(\mathbf{s})}}{\sqrt{s_{10}^{(i)}}\sqrt{p_{U_1}^{(i)}(\mathbf{s})} + \sqrt{s_{20}^{(i)}}\sqrt{p_{U_2}^{(i)}(\mathbf{s})}} \frac{\sqrt{s_{10}^{(i)}}}{\sqrt{s_{20}^{(i)}}} \frac{\sqrt{p_{U_2}^{(i)}(\mathbf{s})}}{\sqrt{p_{U_1}^{(i)}(\mathbf{s})}} = \frac{\lambda_1}{\lambda_2}, \tag{34}$$

which yields

$$p_{U_1}^{(i)}(\mathbf{s}) = \frac{\lambda_2^2 s_{10}^{(i)}}{\lambda_1^2 s_{20}^{(i)}} p_{U_2}^{(i)}(\mathbf{s}). \tag{35}$$

Plugging (35) into (24) (equivalently (27)), we achieve the following crucial equality

$$\frac{\gamma_2 \mu_2}{A^{(i)}} = \frac{\lambda_1 \lambda_2}{\lambda_1 s_{20}^{(i)} + \lambda_2 s_{10}^{(i)}}. \tag{36}$$

The significance of (36) is that, its left hand size, which involves all power components through $A^{(i)}$, and appears in all of (22)-(27), can be replaced by a term which depends only on the fixed Lagrange multipliers, $\lambda_1$ and $\lambda_2$, and the direct link gains, $s_{k0}^{(i)}$. Therefore, the optimality conditions for $p_{1m}^{(i)}(\mathbf{s})$ and $p_{2n}^{(i)}(\mathbf{s})$ can be rewritten independently of $p_{U_k}^{(i)}(\mathbf{s})$. For example, using (36) in (22), we get

$$(\mu_1 - \mu_2 + \gamma_1 \mu_2) \frac{s_{1m}^{(i)}}{1 + s_{1m}^{(i)} p_{1m}^{(i)}(\mathbf{s})} \le \frac{{\lambda_1}^2 s_{20}^{(i)}}{\lambda_1 s_{20}^{(i)} + \lambda_2 s_{10}^{(i)}}, \tag{37}$$

which yields the waterfilling solution, (29a). Similarly, using (36) in (23), (25) and (26), we obtain (30a), (29b) and (30b) respectively. The expression, (28), of optimal $p_{U_1}^{(i)}(\mathrm{s})$ follows from (24), (35) and (36).

Note however that, $p_{U_k}^{(i)}(\mathrm{s})$ obtained by (28) is not guaranteed to be positive. In case it is not, this means that (24) (equivalently (27)) is satisfied with strict inequality, the optimal solution for $p_{U_k}^{(i)}(\mathrm{s})$ should be set to 0 and (36) can no longer be used. Then, when $p_{U_k}^{(i)}(\mathrm{s}) = 0$, instead of (22)-(23) and (25)-(26) we have to apply the conditions:

$$(\mu_1 - \mu_2 + \gamma_1\mu_2)\frac{s_{1m}^{(i)}}{1 + s_{1m}^{(i)}p_{1m}^{(i)}(\mathrm{s})} + \gamma_2\mu_2\frac{s_{10}^{(i)}}{1 + s_{10}^{(i)}p_{1m}^{(i)}(\mathrm{s}) + s_{20}^{(i)}p_{2n}^{(i)}(\mathrm{s})} \leq \lambda_1, \tag{38}$$

$$\gamma_1\mu_2\frac{s_{2n}^{(i)}}{1 + s_{2n}^{(i)}p_{2n}^{(i)}(\mathrm{s})} + \gamma_2\mu_2\frac{s_{20}^{(i)}}{1 + s_{10}^{(i)}p_{1m}^{(i)}(\mathrm{s}) + s_{20}^{(i)}p_{2n}^{(i)}(\mathrm{s})} \leq \lambda_2, \tag{39}$$

for s $\in \mathcal{S}_1 \cup \mathcal{S}_2 \cup \mathcal{S}_3$, and

$$(\mu_1 - \mu_2)\frac{s_{10}^{(i)}}{1 + s_{10}^{(i)}p_{1m}^{(i)}(\mathrm{s})} + \mu_2\frac{s_{10}^{(i)}}{1 + s_{10}^{(i)}p_{1m}^{(i)}(\mathrm{s}) + s_{20}^{(i)}p_{2n}^{(i)}(\mathrm{s})} \leq \lambda_1, \tag{40}$$

$$\mu_2\frac{s_{20}^{(i)}}{1 + s_{20}^{(i)}p_{2n}^{(i)}(\mathrm{s}) + s_{10}^{(i)}p_{1m}^{(i)}(\mathrm{s})} \leq \lambda_2, \tag{41}$$

for s $\in \mathcal{S}_4$.

When $p_{U_k}^{(i)}(\mathrm{s}) = 0$, $k = 1, 2$; the powers $p_{1m}^{(i)}(\mathrm{s})$ and $p_{2n}^{(i)}(\mathrm{s})$ are automatically independent of $p_{U_k}^{(i)}(\mathrm{s})$. However, (38) and (39); (40) and (41) are coupled, and each should be solved by finding the positive roots of a quadratic equation. Since all power values are non-negative, i.e., $p_{1m}^{(i)}(\mathrm{s}) \geq 0$ and $p_{2n}^{(i)}(\mathrm{s}) \geq 0$, we can achieve $p_{1m}^{(i)}(\mathrm{s})$ in (32a), $p_{2n}^{(i)}(\mathrm{s})$ in (33a) by solving (38) and (39). Similarly, $p_{1m}^{(i)}(\mathrm{s})$ in (32b) and $p_{2n}^{(i)}(\mathrm{s})$ in (33b) can be obtained using (40) and (41). $\gamma_1$, $\lambda_1$ and $\lambda_2$ are selected in such a way that, when the power levels in (28)-(33b) are used, the constraints (15)-(18) are satisfied. ■

The power levels of the cooperative codewords on each subchannel, $p_{1m}^{(i)}(\mathrm{s})$ and $p_{2n}^{(i)}(\mathrm{s})$ in (29a) and (30a), have an interesting single user waterfilling type interpretation, as they solely depend on the channel gains of only that particular subchannel, and the fixed Lagrange multipliers. The water level is determined by the direct link gains. However, in (32a) and (33a) the power $p_{1m}^{(i)}(\mathrm{s})$ depends on $p_{2n}^{(i)}(\mathrm{s})$, and vice-versa: increasing one of the powers will decrease the other, should the constraints (38)-(41) be satisfied with equality, and we now have a multi-user waterfilling type solution. This is somewhat different than the observations in [8], which conjectured that a single user waterfilling type solution for cooperative powers would be sufficient in all scenarios, for the much simpler case of the scalar MAC, and sum rate maximization only.

At this point, it should be clear that although (29a)-(30b) and (32a)-(33b) do not explicitly depend on $p_{U_k}^{(i)}(\mathbf{s})$, the decision regarding which of these equations should be used while computing $p_{kj}^{(i)}(\mathbf{s})$ does. Likewise, $p_{U_k}^{(i)}(\mathbf{s})$ are clearly functions of $p_{kj}^{(i)}(\mathbf{s})$, which makes equations (29a)-(30b), (32a)-(33b) and (38)-(41) coupled. Note however that, the way we proved Theorem 1 automatically suggests a natural way of solving the KKT conditions iteratively. To this end, we propose an algorithm which performs updates on the powers of the users, one-user-at-a-time: given $p_{U_1}^{(i)}(\mathbf{s})$ and $p_{12}^{(i)}(\mathbf{s})$, it computes $p_{U_2}^{(i)}(\mathbf{s})$ and $p_{21}^{(i)}(\mathbf{s})$, and using these new values for user 2, it re-iterates the powers of user 1. This algorithm simplifies the seemingly difficult task of obtaining the optimal powers from the coupled equations, and due to the convex nature of the problem, and the Cartesian nature of the constraints across users, it provably converges to the optimal solution, as at the end of the iterations, the KKT conditions will be satisfied. The outline of the algorithm is given below.

---

**Algorithm 1** Iterative Power Allocation Algorithm

---

**for** $\mu_2 = 0 : 1$ **do**
  **while** (15)-(16) are not satisfied **do**
    **while** (17) is not satisfied **do**
      Calculate $p_{1m}^{(i)}(\mathbf{s})$ using (29a)-(29b) and $p_{U_1}^{(i)}(\mathbf{s})$ using (28) assuming $p_{U_1}^{(i)}(\mathbf{s}) > 0$, $\forall i$
      **while** $\exists$ s′ s.t. $p_{U_1}^{(i)}(\mathrm{s}') < 0$ **do**
        Set $p_{U_1}^{(i)}(\mathrm{s}') = 0$ and re-calculate $p_{1m}^{(i)}(\mathrm{s}')$ using (32a)-(32b) and $p_{U_1}^{(i)}(\mathrm{s}')$ using (28)
      **end while**
      Update $\lambda_1$
    **end while**
    **while** (18) is not satisfied **do**
      Calculate $p_{2n}^{(i)}(\mathbf{s})$ using (30a)-(30b) and $p_{U_2}^{(i)}(\mathbf{s})$ using (28) assuming $p_{U_2}^{(i)}(\mathbf{s}) > 0$, $\forall i$
      **while** $\exists$ s′ s.t. $p_{U_2}^{(i)}(\mathrm{s}') < 0$ **do**
        Set $p_{U_2}^{(i)}(\mathrm{s}') = 0$ and re-calculate $p_{2n}^{(i)}(\mathrm{s}')$ using (33a)-(33b) and $p_{U_2}^{(i)}(\mathrm{s}')$ using (28)
      **end while**
      Update $\lambda_2$
    **end while**
    Update $\gamma_1$
  **end while**
**end for**

---

Perhaps the most important feature of this algorithm is that, *regardless of the number of subchannels used*, we only need to solve for three Lagrange multipliers, which relate the powers allocated to the subchannels, to obtain the optimum PA. This reduces the complexity of the algorithm dramatically, and makes it scalable, compared to the subgradient algorithm. As a result, the convergence is much faster.

## V. Simulation Results

In order to obtain the optimal PA policy, and the resulting achievable rate region, we implement the projected subgradient algorithm, and the iterative waterfilling-like algorithm based on Karush-Kuhn-Tucker (KKT) conditions on optimality, for a simple case with only three subchannels. The achievable rate region for the InterSCE strategy is obtained by running this algorithm for varying priorities $\mu_k$, and then by taking a convex hull over the resulting power optimized regions. In Figure 2, we compare the achievable rate regions for power controlled cooperative OFDMA utilizing the projected subgradient algorithm and the iterative algorithm, with those for several encoding strategies without power control, from [13]. We assume that, for the channel non-adaptive protocols, the users are still able to allocate their total power across subchannels and codewords. The total power of each user and the noise variances are set to unity. The fading coefficients are chosen from independent Rayleigh distributions, the means of which are shown in Figure 2. We observe that, when the powers are chosen jointly optimally with InterSCE, there is a major improvement in achievable rates. This unusually high gain from power control can be attributed to our ability to take advantage of the additional diversity created by OFDMA: PA not only allows us to use the subchannels at time varying instantaneous rates based on the channel qualities, but also to use them adaptively for varying purposes, i.e., cooperation, common message generation or direct transmission.

In practice, the feedback channel from the receiver to the transmitters can send only a few bits of feedback, as otherwise a significant portion of channel resources have to be allocated to the reverse link which does not contribute to the channel rate. Hence, in Figure 2, we also show the rate region achievable with limited feedback. We assume that, since the feedback is very low rate, it is error free. When the receiver has access to channel gains of the users, there are two approaches one can take to feed back information to the users: a straightforward method is to quantize the channel states, and feed back the quantized fading values on each subchannel. Assuming a $Q$ bit quantizer is used for each fading state, the receiver should feed back a total of $4NQ$ bits of CSI to each user, and then the users will have to look up the power levels optimized for the quantized channel states, and use them in their transmission. An alternative method is to compute the optimal power levels first at the receiver, and then quantize them to obtain a quantized power codebook. Whenever a channel state is observed, the receiver can then directly feed back the quantized powers to be used to the users. Due to the structure of the optimal PA policy observed in Lemma 1, only two powers out of three are active for each user $k$ at any given channel state, and which one will be active only depends on a single comparison, $s_{kj}^{(i)} \lessgtr s_{k0}^{(i)}$, which requires only one bit feedback. Hence, the total feedback required per user is $(2Q+1)N$ bits per user, assuming $Q$ bit feedback is used for each power value. We use Lloyd-Max algorithm to quantize the powers, taking into account their probability distribution induced by the underlying channel state distribution. The case with $Q = 1$ is plotted on Figure 2. A quite interesting observation is that even with one bit feedback per power component, which is equivalent to selecting one of two possible values for each codeword's power, a large improvement in rates can be achieved compared to the non power-controlled scenario. With two bits of feedback per component, the achievable rate region is nearly the same as that for perfect CSI, and is omitted to avoid confusion with the subgradient rate region.

In Figure 2, it is also observed that the gain achieved by power control through the iterative algorithm always exceeds the projected subgradient algorithm, especially in the sum rate region. The main reason is that, the subgradient algorithm had still not fully converged, when it was stopped at 10000 iterations, while the iterative algorithm did fully converge to the optimal PA. The relative convergence times of the two algorithms are shown in Figure 3, which clearly depicts the advantage of using the iterative algorithm over the subgradient algorithm.

In Figure 4, we compare the rate regions in a uniform fading environment with means expressed on the figure. Here we ensure $\mathrm{s} \in \mathcal{S}_1$, with the motivation of obtaining a strictly optimal PA, and a simpler description of the power distributions. In this setting, since some of the power values are always zero, the number of power variables is less, and hence the subgradient algorithm nearly converges to the optimum within 10000 iterations, and the rate regions of subgradient and iterative algorithms nearly coincide. For this setting, we further analyze the optimal power distributions over the channel states, in Figures 5(a)-5(c), 6(a)-6(c) and 7(a)-7(b).

Figures 5(a)-5(c) and 6(a)-6(c) demonstrate the optimal powers allocated to subchannel 1, as functions of the inter-user link gains, when the direct link gains are fixed to two different sets specified on the figures. Powers $p_{U_2}^{(1)}$ are not shown, to save space, as they are identical to $p_{U_1}^{(1)}$ due to the symmetry in fading. In Figures 5(a)-5(c), the direct link gains are at their maximum, hence the cooperative powers, $p_{U_k}^{(1)}$, are always positive. In this case, we observe the expected single user waterfilling type behavior for the distributions of $p_{12}^{(1)}(\mathbf{s})$ and $p_{21}^{(1)}(\mathbf{s})$. In Figures 6(a)-6(c) however, when the direct links are moderate on the average, we have a more interesting scenario: when $s_{21}^{(1)}$ is significantly stronger instantaneously, only user 2 uses the subchannel. When both inter-user links are instantaneously strong, the users exchange information using simultaneous waterfilling, and set $p_{U_k}^{(1)}$ to zero. When both inter-user links are weak, the users use the subchannel solely to convey common information to the RX, by using only $p_{U_1}^{(1)}$ and $p_{U_2}^{(1)}$. An important observation is that, although we make no prior assumptions on subchannel allocation to users/codewords, the optimal powers sometimes dictate exclusive use of some subchannels for dedicated tasks. The resulting power distributions show that the KKT conditions are indeed satisfied at the fixed point of our iterative algorithm, verifying convergence.

In Figures 7(a)-7(b), we plot the power distributions obtained using the subgradient algorithm instead, for the same setting as in Figures 6(a)-6(c). The subgradient algorithm is terminated after 10000 iterations. It is observed that while the powers $p_{12}^{(1)}(\mathbf{s})$ and $p_{21}^{(1)}(\mathbf{s})$ seem to have nearly converged to the optimal values shown in Figures 6(a)-6(c) (only $p_{21}^{(1)}(\mathbf{s})$ is shown, as $p_{12}^{(1)}(\mathbf{s})$ is simply symmetrical), the cooperative power $p_{U_1}^{(1)}(\mathbf{s})$ has still not fully converged, though it is close to its optimal distribution. Note that, the effect of this is negligible on the rate regions, as was shown in Figure 4.

## VI. CONCLUSION

We obtained the optimum PA policies for a mutually cooperative OFDMA channel employing IntraSCE and InterSCE strategies. We developed a subgradient algorithm and a more efficient iterative algorithm which maximize the achievable rate region. The number of iterations of the iterative algorithm does not depend on the number of subchannels, which makes the algorithm scalable. We demonstrated that the optimal PA may also serve as a guideline for subchannel assignment to the users' cooperative codewords, and that PA for cooperative OFDMA provides significant rate improvements, even in limited feedback scenarios, due to its ability to exploit the diversity provided by OFDMA.

## VII. APPENDIX

Note that KKT conditions are necessary and sufficient for optimality. To obtain the KKT conditions we first assign the Lagrange multipliers $\gamma_1$, $\gamma_2$, $\lambda_1$ and $\lambda_2$ to the inequality constraints (15), (16), (17), (18) respectively, and we further assign $\epsilon_t^i(\mathbf{s})$, $t = 1, \ldots, 6$, $\forall$s to the positivity constraints (19), to obtain the Lagrangian

$$
\begin{aligned}
\mathcal{L} = R_\mu + \gamma_1 \Bigg[ (\mu_1 - \mu_2) \left( \sum_{i=1}^{N} E_{S_1,S_2} \left[ C(p_{12}^{(i)}(\mathbf{s}) s_{12}^{(i)}) \right] + \sum_{i=1}^{N} E_{S_3,S_4} \left[ C(p_{10}^{(i)}(\mathbf{s}) s_{10}^{(i)}) \right] \right) \\
+ \mu_2 \sum_{i=1}^{N} \Bigg( E_{S_1} \left[ C(p_{12}^{(i)}(\mathbf{s}) s_{12}^{(i)}) + C(p_{21}^{(i)}(\mathbf{s}) s_{21}^{(i)}) \right] + E_{S_2} \left[ C(p_{12}^{(i)}(\mathbf{s}) s_{12}^{(i)}) + C(p_{20}^{(i)}(\mathbf{s}) s_{20}^{(i)}) \right] \\
+ E_{S_3} \left[ C(p_{10}^{(i)}(\mathbf{s}) s_{10}^{(i)}) + C(p_{21}^{(i)}(\mathbf{s}) s_{21}^{(i)}) \right] + E_{S_4} \left[ C(p_{10}^{(i)}(\mathbf{s}) s_{10}^{(i)} + p_{20}^{(i)}(\mathbf{s}) s_{20}^{(i)}) \right] \Bigg) - R_\mu \Bigg] \\
+ \gamma_2 \Bigg[ (\mu_1 - \mu_2) \left( \sum_{i=1}^{N} E \left[ C(p_{1m}^{(i)}(\mathbf{s}) s_{1m}^{(i)}) \right] \right) + \mu_2 \sum_{i=1}^{N} E \Big[ C \Big( s_{10}^{(i)} (p_{1m}^{(i)}(\mathbf{s}) + p_{U_1}^{(i)}(\mathbf{s})) \\
+ s_{20}^{(i)} (p_{2n}^{(i)}(\mathbf{s}) + p_{U_2}^{(i)}(\mathbf{s})) + 2\sqrt{s_{10}^{(i)} s_{20}^{(i)} p_{U_1}^{(i)}(\mathbf{s}) p_{U_2}^{(i)}(\mathbf{s})} \Big) \Big] - R_\mu \Bigg] \\
+ \lambda_1 \left( \bar{p}_1 - \sum_{i=1}^{N} \left( E_{S_3,S_4} \left[ p_{10}^{(i)}(\mathbf{s}) \right] + E_{S_1,S_2} \left[ p_{12}^{(i)}(\mathbf{s}) \right] + E \left[ p_{U_1}^{(i)}(\mathbf{s}) \right] \right) \right) \\
+ \lambda_2 \left( \bar{p}_2 - \sum_{i=1}^{N} \left( E_{S_2,S_4} \left[ p_{20}^{(i)}(\mathbf{s}) \right] + E_{S_1,S_3} \left[ p_{21}^{(i)}(\mathbf{s}) \right] + E \left[ p_{U_2}^{(i)}(\mathbf{s}) \right] \right) \right) \\
+ \epsilon_1^{(i)}(\mathbf{s}) p_{10}^{(i)}(\mathbf{s}) + \epsilon_2^{(i)}(\mathbf{s}) p_{12}^{(i)}(\mathbf{s}) + \epsilon_3^{(i)}(\mathbf{s}) p_{U_1}^{(i)}(\mathbf{s}) + \epsilon_4^{(i)}(\mathbf{s}) p_{20}^{(i)}(\mathbf{s}) + \epsilon_5^{(i)}(\mathbf{s}) p_{21}^{(i)}(\mathbf{s}) + \epsilon_6^{(i)}(\mathbf{s}) p_{U_2}^{(i)}(\mathbf{s}). \quad (42)
\end{aligned}
$$

For s $\in \mathcal{S}_1 \cup \mathcal{S}_2 \cup \mathcal{S}_3$, we take partial derivatives of the Lagrangian function, $\mathcal{L}$ with respect to $p_{1m}^{(i)}(\mathrm{s})$, $p_{2n}^{(i)}(\mathrm{s})$, and $p_{U_k}^{(i)}(\mathrm{s})$, $\forall i$ and $\forall$s, to obtain the respective conditions

$$\gamma_2\mu_2\left(\frac{s_{10}^{(i)}}{A^{(i)}}\right) + (\mu_1 - \mu_2 + \gamma_1\mu_2)\left(\frac{s_{1m}^{(i)}}{1 + s_{1m}^{(i)}p_{1m}^{(i)}(\mathrm{s})}\right) - \lambda_1 + \epsilon_{e_1}^{(i)}(\mathrm{s}) = 0, \tag{43}$$

$$\gamma_2\mu_2\left(\frac{s_{20}^{(i)}}{A^{(i)}}\right) + \gamma_1\mu_2\left(\frac{s_{2n}^{(i)}}{1 + s_{2n}^{(i)}p_{2n}^{(i)}(\mathrm{s})}\right) - \lambda_2 + \epsilon_{e_2}^{(i)}(\mathrm{s}) = 0, \tag{44}$$

$$\gamma_2\mu_2\left(\frac{\sqrt{s_{k0}^{(i)}s_{j0}^{(i)}p_{U_j}^{(i)}(\mathrm{s})} + s_{k0}^{(i)}\sqrt{p_{U_k}^{(i)}(\mathrm{s})}}{A^{(i)}\sqrt{p_{U_k}^{(i)}(\mathrm{s})}}\right) - \lambda_k + \epsilon_{e_3}^{(i)}(\mathrm{s}) = 0, \tag{45}$$

where $e_1 \in \{1, 2\}$, $e_2 \in \{4, 5\}$ and $e_3 \in \{3, 6\}$ take their values based on with respect to which power the derivative is taken. Likewise, for s $\in \mathcal{S}_4$, and the respective partial derivatives yield

$$\gamma_2\mu_2\left(\frac{s_{10}^{(i)}}{A^{(i)}}\right) + (\mu_1 - \mu_2)\left(\frac{s_{10}^{(i)}}{1 + s_{10}^{(i)}p_{1m}^{(i)}(\mathrm{s})}\right)$$
$$+\gamma_1\mu_2\left(\frac{s_{10}^{(i)}}{1 + s_{10}^{(i)}p_{1m}^{(i)}(\mathrm{s}) + s_{20}^{(i)}p_{2n}^{(i)}(\mathrm{s})}\right) - \lambda_1 + \epsilon_{e_1}^{(i)}(\mathrm{s}) = 0, \tag{46}$$

$$\gamma_2\mu_2\left(\frac{s_{20}^{(i)}}{A^{(i)}}\right) + \gamma_1\mu_2\left(\frac{s_{20}^{(i)}}{1 + s_{10}^{(i)}p_{1m}^{(i)}(\mathrm{s}) + s_{2n}^{(i)}p_{20}^{(i)}(\mathrm{s})}\right) - \lambda_2 + \epsilon_{e_2}^{(i)}(\mathrm{s}) = 0, \tag{47}$$

$$\gamma_2\mu_2\left(\frac{\sqrt{s_{k0}^{(i)}s_{j0}^{(i)}p_{U_j}^{(i)}(\mathrm{s})} + s_{k0}^{(i)}\sqrt{p_{U_k}^{(i)}(\mathrm{s})}}{A^{(i)}\sqrt{p_{U_k}^{(i)}(\mathrm{s})}}\right) - \lambda_k + \epsilon_{e_3}^{(i)}(\mathrm{s}) = 0. \tag{48}$$

Since the optimal PA policy should satisfy the complementary slackness constraints,

$$p_{10}^{(i)}(\mathrm{s})\epsilon_1^{(i)}(\mathrm{s}) = 0, \quad p_{12}^{(i)}(\mathrm{s})\epsilon_2^{(i)}(\mathrm{s}) = 0, \quad p_{U_1}^{(i)}(\mathrm{s})\epsilon_3^{(i)}(\mathrm{s}) = 0,$$
$$p_{20}^{(i)}(\mathrm{s})\epsilon_4^{(i)}(\mathrm{s}) = 0, \quad p_{21}^{(i)}(\mathrm{s})\epsilon_5^{(i)}(\mathrm{s}) = 0, \quad p_{U_2}^{(i)}(\mathrm{s})\epsilon_6^{(i)}(\mathrm{s}) = 0, \tag{49}$$

we can either drop $\epsilon_t^{(i)}(\mathrm{s})$ in each of (43)-(48), if the corresponding power is positive; or we can replace the equality by a strict inequality, meaning that $\epsilon_t^{(i)}(\mathrm{s})$ is non-zero but its corresponding power is zero. Hence, using the relevant conditions from (49) in (43)-(48), and dropping the dependencies on $\epsilon_t^{(i)}(\mathrm{s})$, we write the conditions for optimality in terms of inequalities instead, which yield (22)-(27). The inequalities hold with equality if and only if the corresponding power level is positive, and with strict inequality of that power level is zero.

Partial derivatives with respect to the dual variables dictate that the conditions (15)-(18) are satisfied. Finally, partial derivatives with respect to $R_\mu$ yields $\gamma_1 + \gamma_2 = 1$, hence the condition $\gamma_1 = 1 - \gamma_2$.

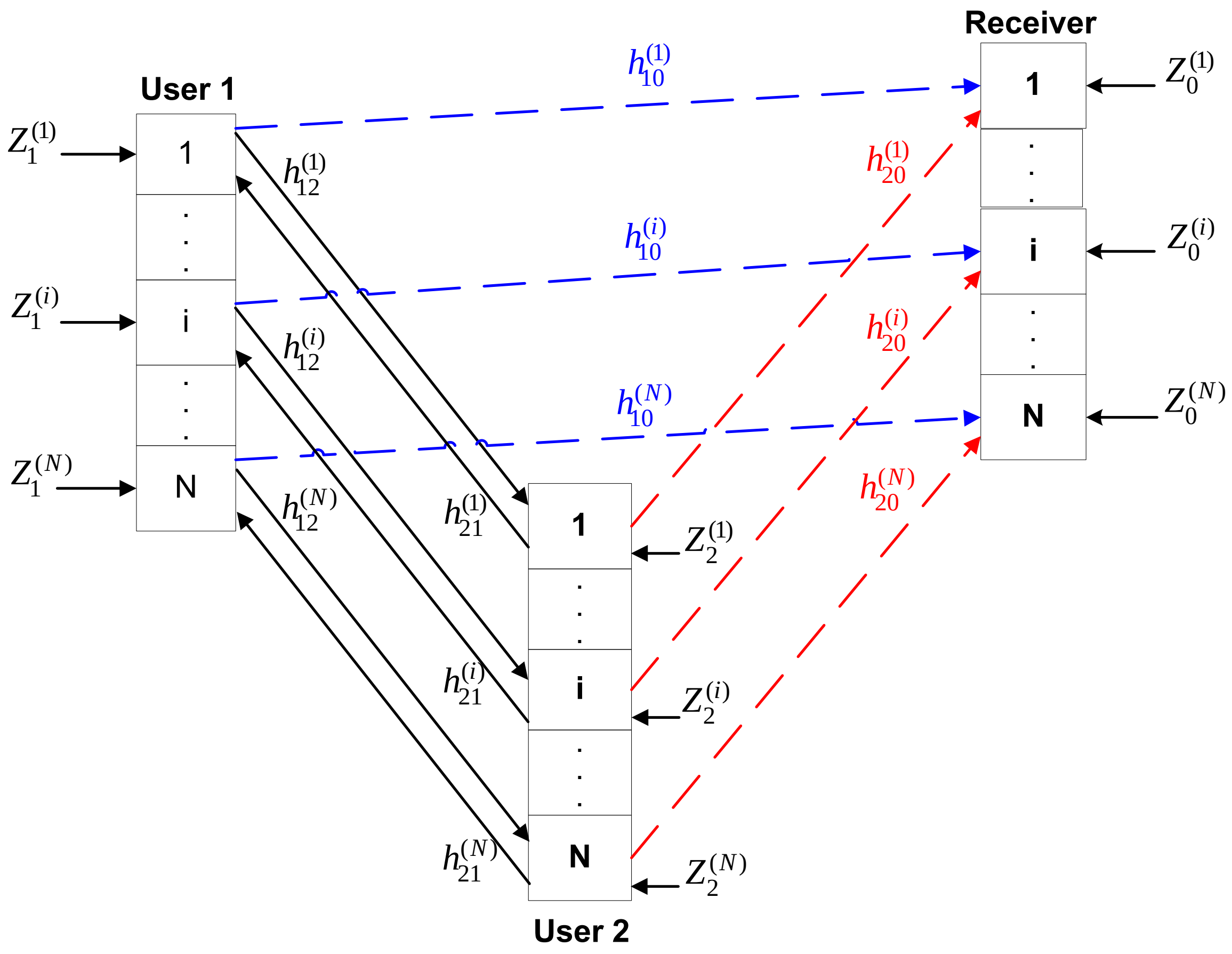

Fig. 1.   Gaussian cooperative OFDMA channel.

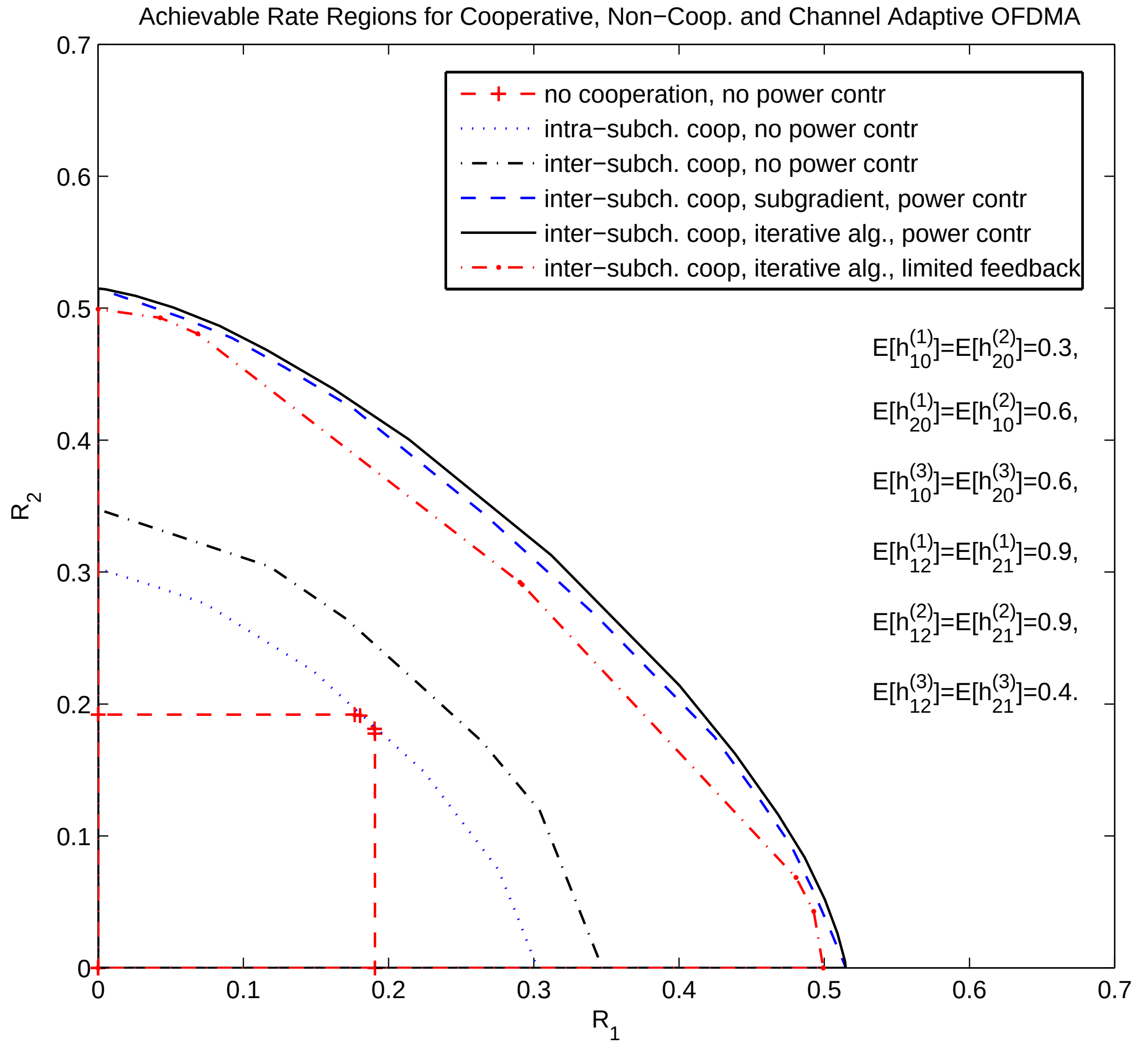

Fig. 2. Achievable rate regions in Rayleigh fading.

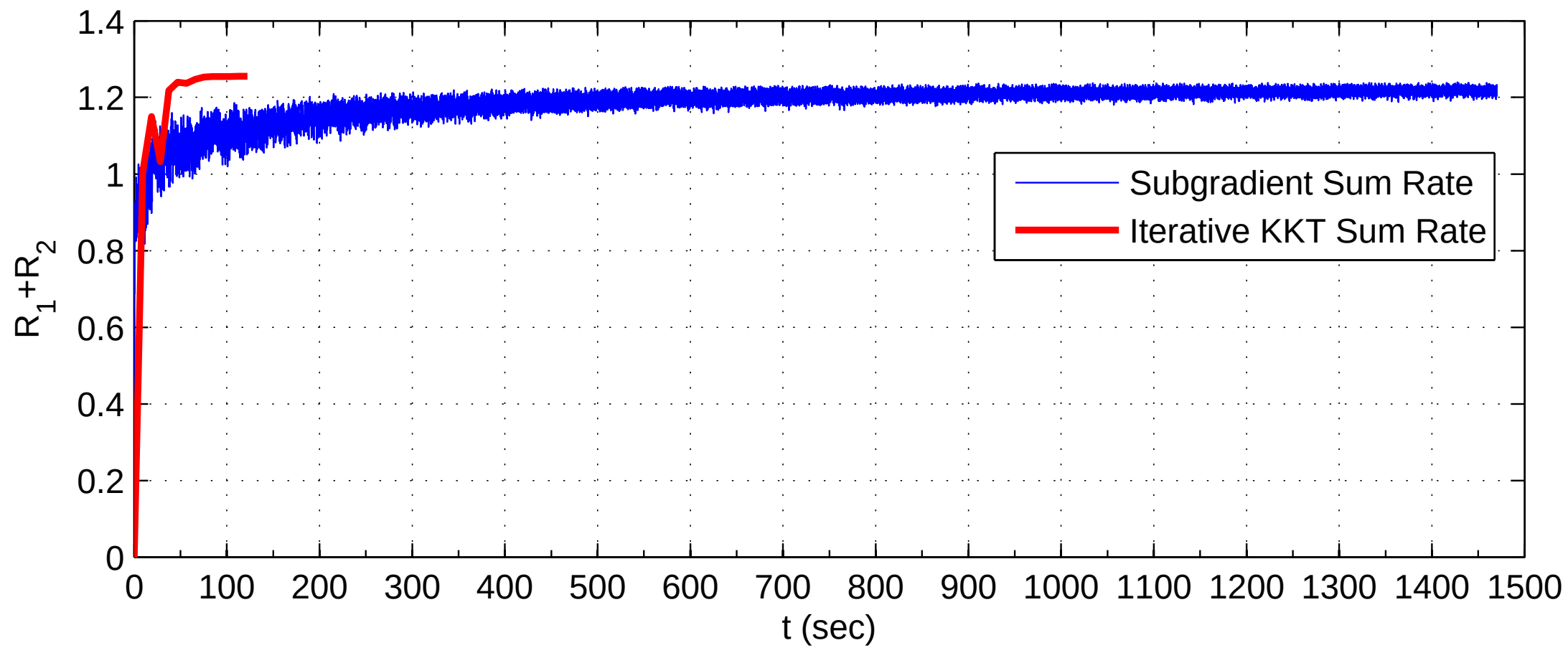

Fig. 3. Comparison of the convergence times of the proposed algorithms in Rayleigh fading.

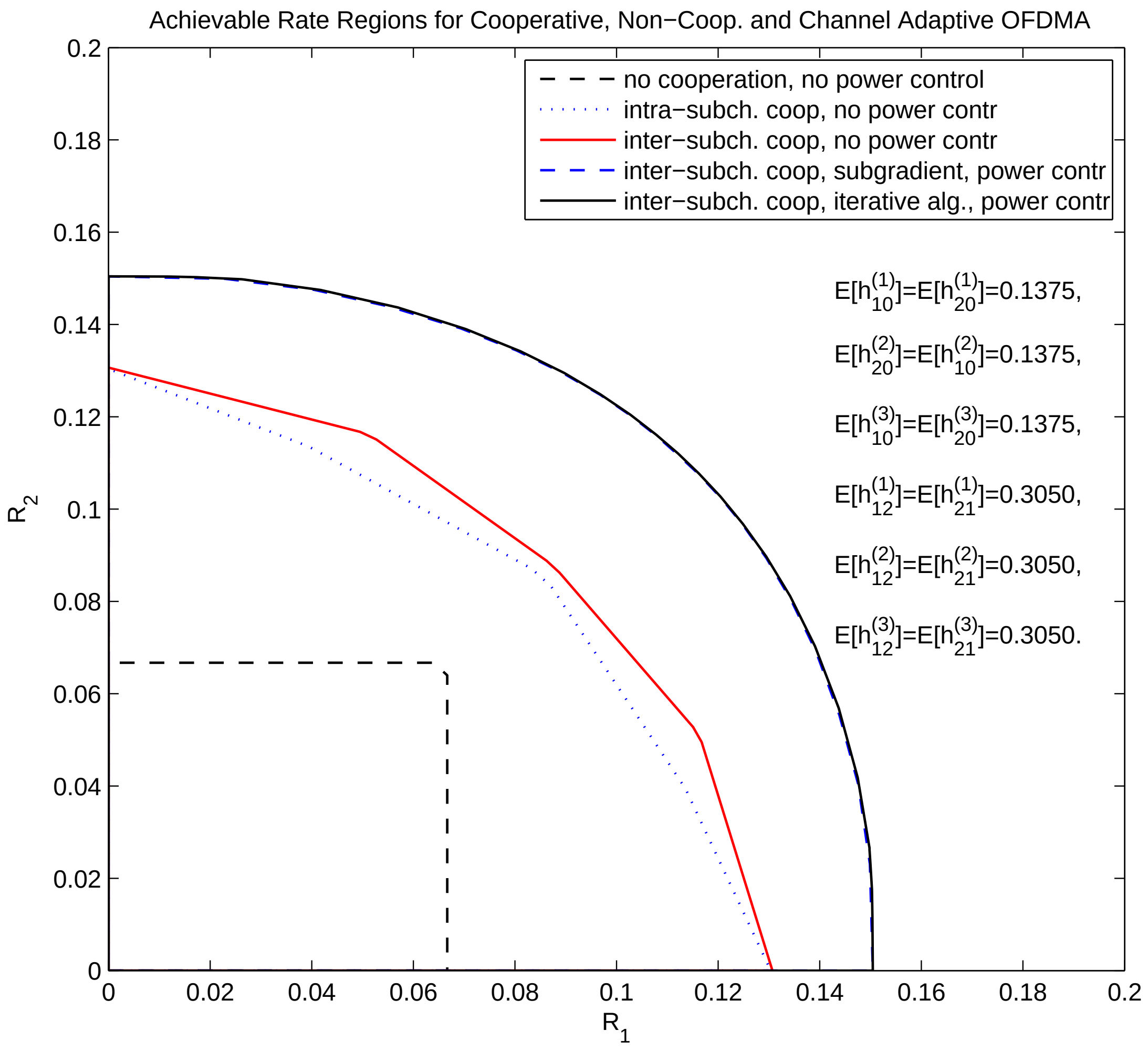

Fig. 4. Achievable rate regions in uniform fading.

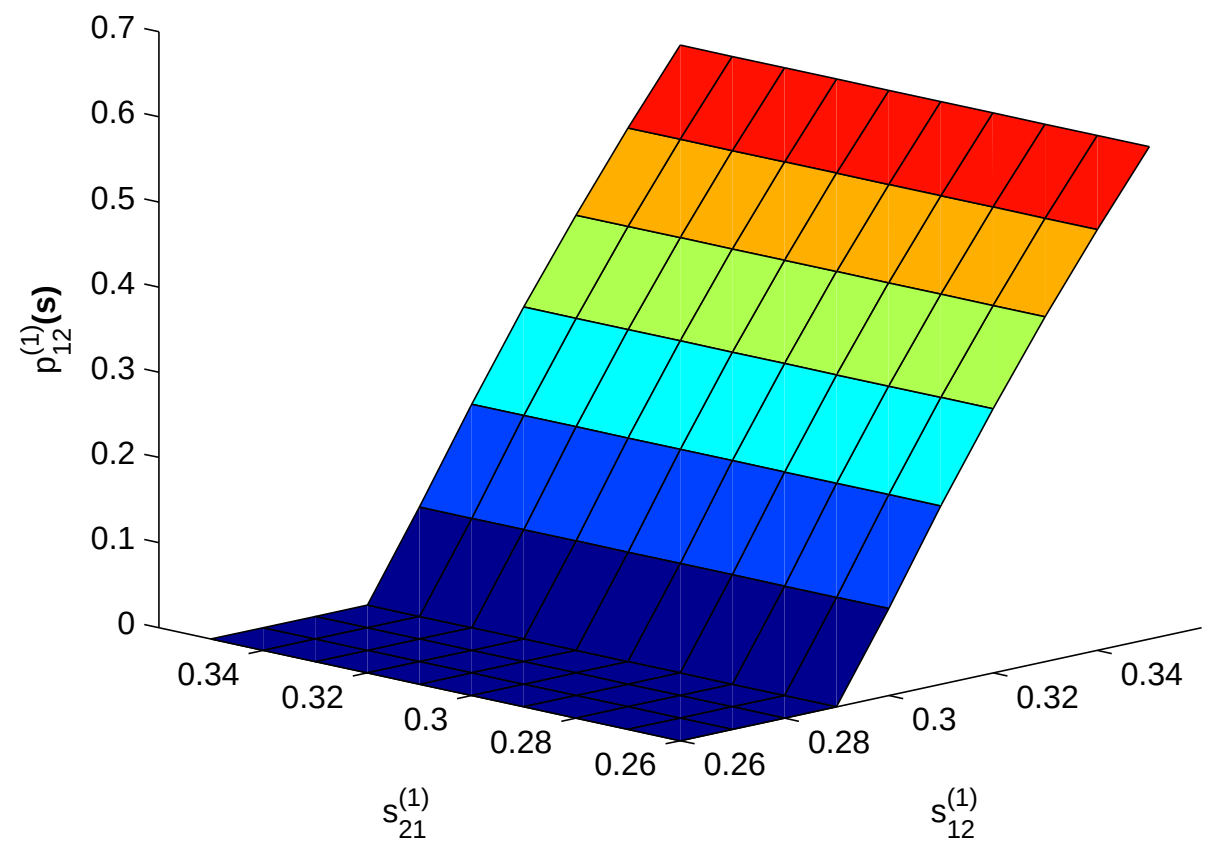

(a) Power level, $p_{12}^{(1)}$

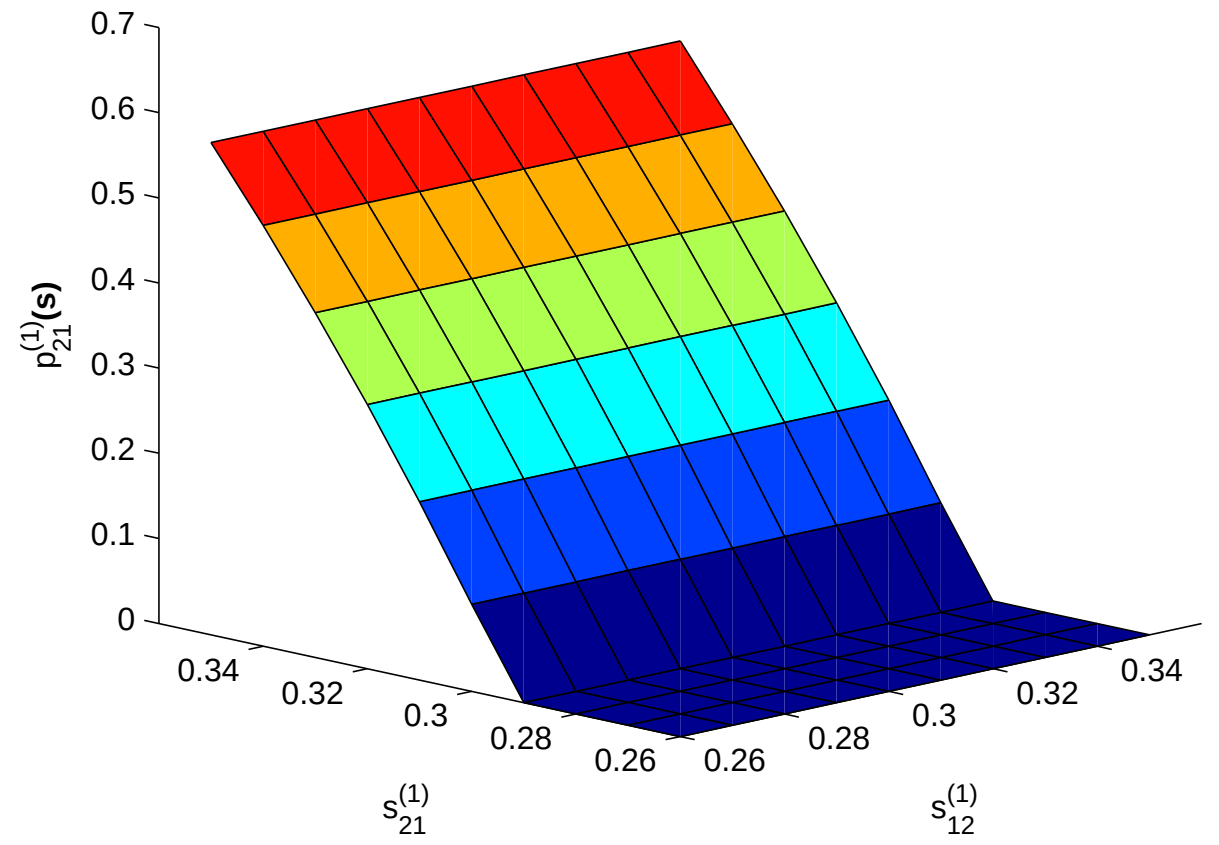

(b) Power level, $p_{21}^{(1)}$

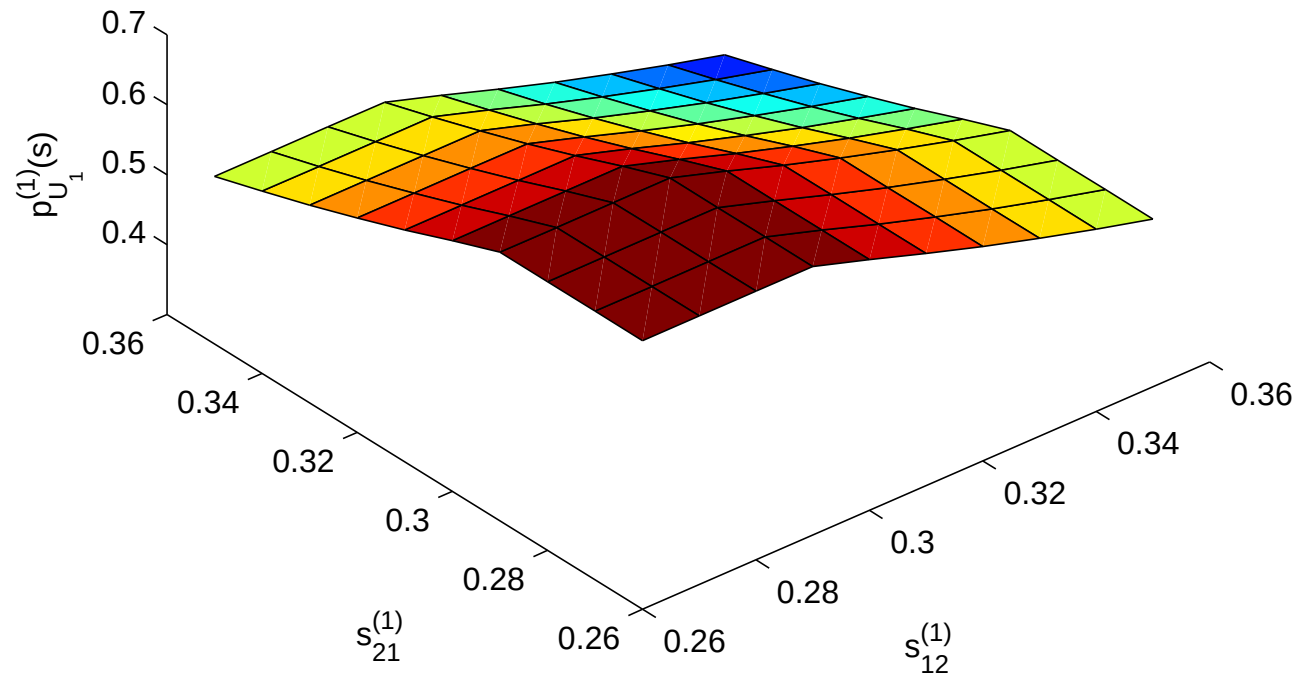

(c) Power level, $p_{U1}^{(1)} = p_{U2}^{(1)}$

Fig. 5. Optimal power allocation when $s_{10}^{(1)}$ and $s_{20}^{(1)}$ are maximum ($i.e.$ $s_{10}^{(1)} = s_{20}^{(1)} = 0.25$), fixed and always less than $s_{12}^{(1)}$ and $s_{21}^{(1)}$. $p_{U_k}^{(1)}$ are always positive, to take advantage of strong direct links. $p_{kj}^{(1)}$ obey single user waterfilling, as expected.

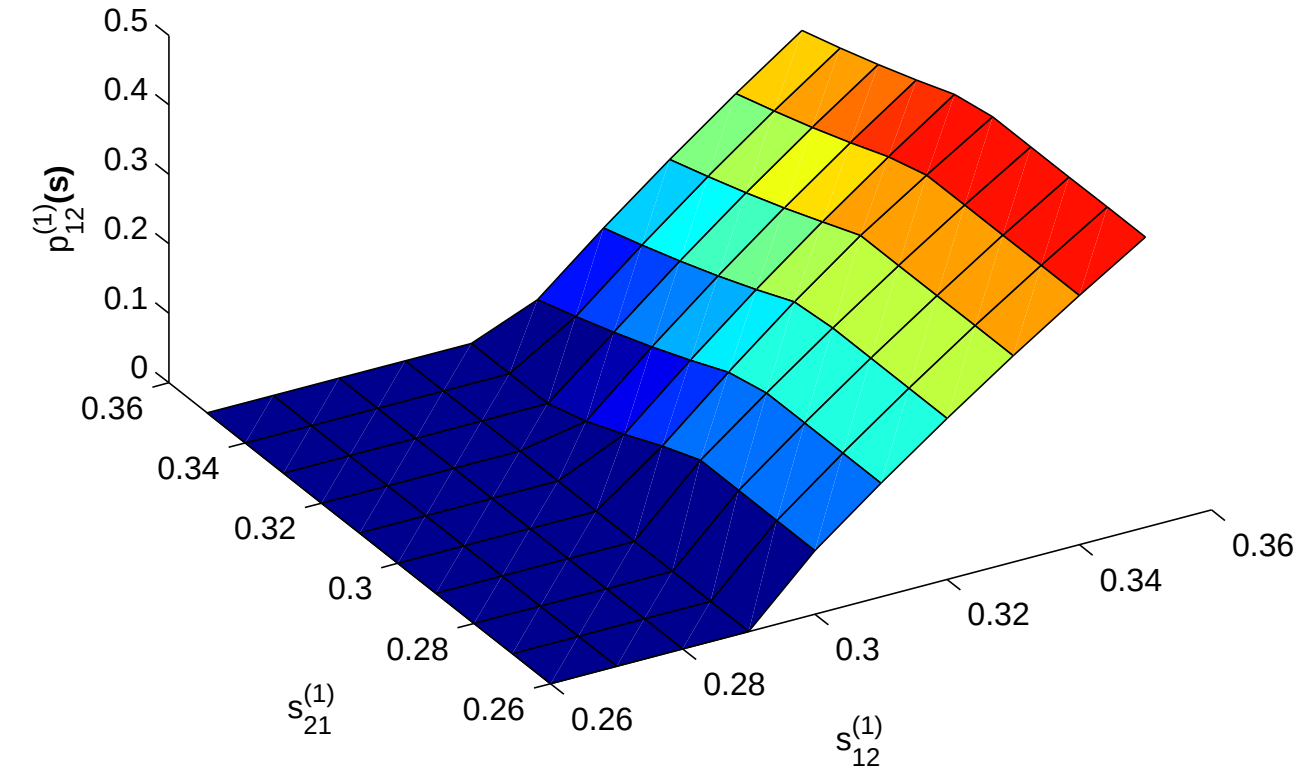

(a) Power level, $p_{12}^{(1)}$

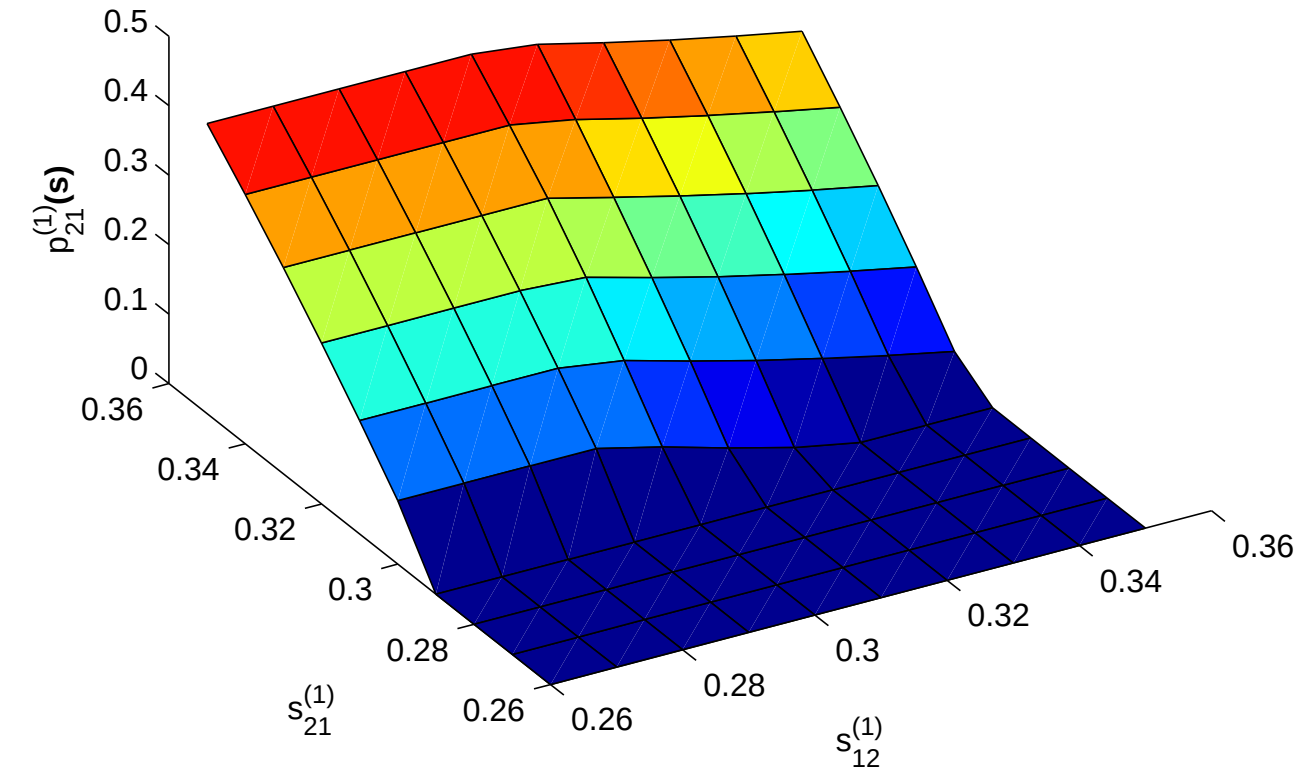

(b) Power level, $p_{21}^{(1)}$

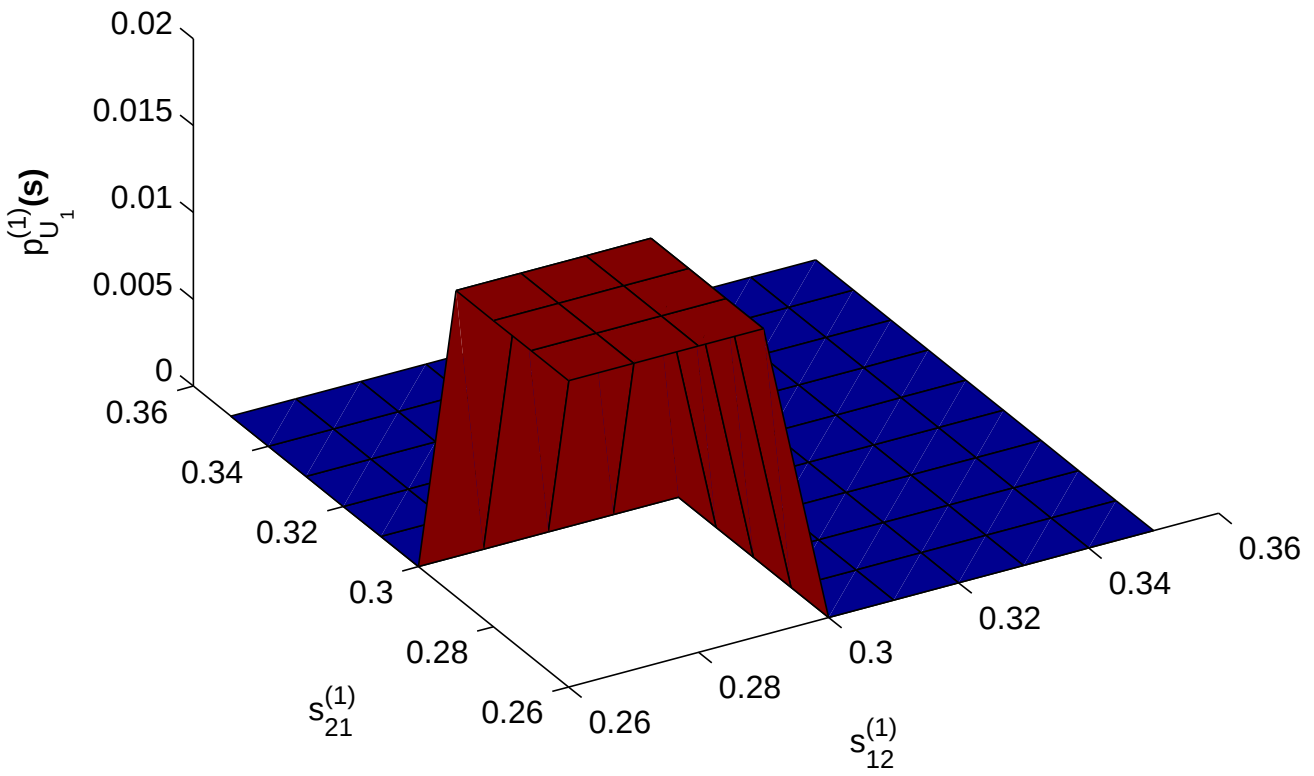

(c) Power level, $p_{U1}^{(1)} = p_{U2}^{(1)}$

Fig. 6. Optimal power allocation when $s_{10}^{(1)} = s_{20}^{(1)} = 0.15$, fixed and always less than $s_{12}^{(1)}$ and $s_{21}^{(1)}$. When $p_{U_k}^{(1)}$ is positive, $p_{kj}^{(1)}$ obey single user waterfilling. As the inter-user links get stronger, it becomes more profitable to create common information, $p_{U_k}^{(1)}$ become 0, and the users perform simultaneous waterfilling.

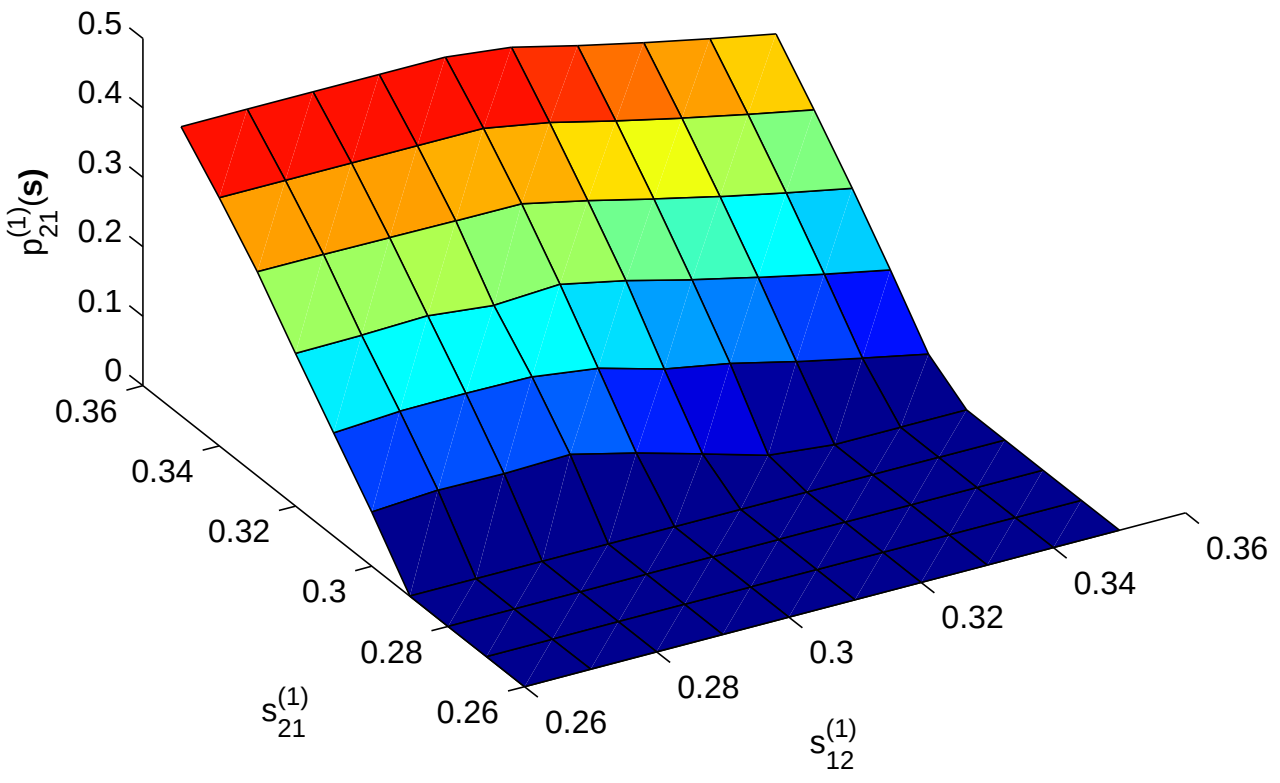

(a) Power level, $p_{21}^{(1)}$

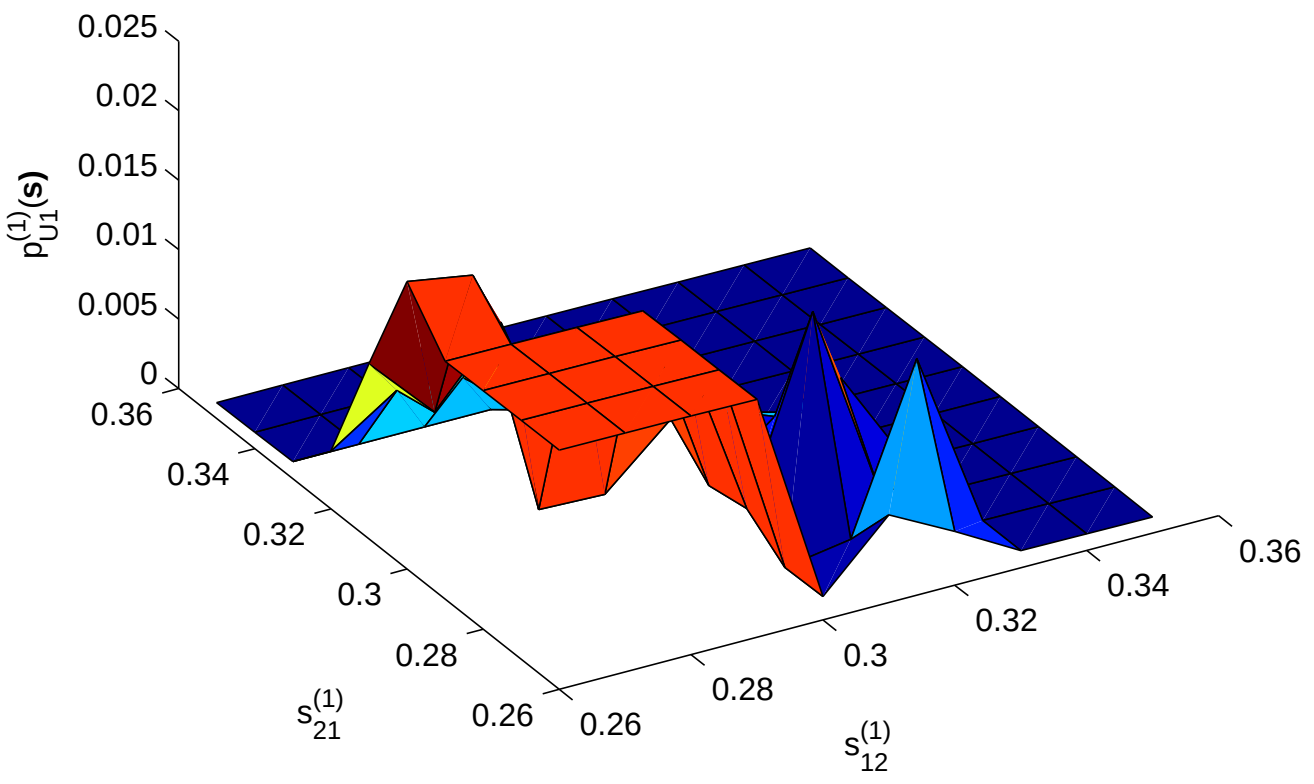

(b) Power level, $p_{U1}^{(1)}$

Fig. 7. Power allocation obtained after 10000 iterations of the subgradient algorithm, when $s_{10}^{(1)} = s_{20}^{(1)} = 0.15$, fixed and always less than $s_{12}^{(1)}$ and $s_{21}^{(1)}$. The algorithm has not yet converged to the optimum value, despite a much longer running time compared to the iterative algorithm. Achievable rates are nearly within 0.1% of the optimum value.